\documentclass[final,1p, times]{elsarticle}

\usepackage{amssymb}
\usepackage{graphics}
\usepackage{esvect}
\usepackage{multicol}
\usepackage{graphicx}
\usepackage{dcolumn}
\usepackage{bm}
\usepackage{booktabs}
\usepackage{hyperref}
\usepackage{subfigure}
\usepackage{multirow}
\usepackage{float}

\begin{document}

\begin{frontmatter}



\title{Universal scaling of meson and baryon spectra in p-Pb collisions at 5.02 TeV}

\author{Na Liu, Xiaoling Du, Liyun Qiao, Guorong Che}
\author{Wenchao Zhang\corref{cor1}}
\cortext[cor1]{Corresponding author}
\ead{wenchao.zhang@snnu.edu.cn}
\address{School of Physics and Information Technology, Shaanxi Normal University, Xi'an 710119, P. R. China}

\begin{abstract}
We systematically investigate the scaling property of mesons (pions and kaons) and baryons (protons, ${\rm \Lambda}$, ${\rm \Xi}$ and ${\rm \Omega}$) transverse momentum ($p_{\rm T}$) spectra at different centrality classes (0-5$\%$, 5-10$\%$, 10-20$\%$, 20-40$\%$, 40-60$\%$, 60-80$\%$ and 80-100$\%$) in proton-lead collisions with center of mass energy per nucleon pair 5.02 TeV.  In the low $p_{\rm T}$ region with $p_{\rm{T}}\leq $ 3.9 (3.1, 2.5, 2.7, 2.4 and 2.8) GeV/c, a universal scaling independent of the centrality is observed in the pion (kaon, proton, ${\rm \Lambda}$, ${\rm \Xi}$ and ${\rm \Omega}$) spectra when a dilatation, $p_{\rm T}\rightarrow p_{\rm T}/K$, is applied. Here $K$ is a scaling parameter depending on the centrality class. We find that the rates at which ln$K$ changes with the logarithmic value of the average value of the number of participating nucleons, ln$\langle N_{\rm{part}}\rangle$, are stronger for baryons than those for mesons. In the high $p_{\rm T}$ region, there is a deviation from the scaling. The more peripheral the collisions are, the more obvious the violation of the scaling is. In the framework of the colour string percolation (CSP) model, we show that mesons and baryons are generated from the decay of clusters formed by strings overlapping in the transverse plane with the same size dispersion but with different mean size. The mean size of clusters for baryons is smaller than that of mesons. For the same hadrons at different centrality classes, the mean size of clusters decreases with the increase of centrality. The fragmentation functions for cluster decay are different for different hadrons, while they are universal for the same hadrons at different centrality classes. The universal scaling of the meson and baryon spectra in the low $p_{\rm T}$ region can be quantitatively understood with the CSP model at the same time.
\end{abstract}
 
\begin{keyword}
proton-lead collisions\sep $p_{\rm T}$ spectra\sep universal scaling\sep colour string percolation



\end{keyword}

\end{frontmatter}


\section{Introduction}
\label{sec:intro}
Investigating the dynamics of particle productions is one of the main goals in high energy collisions. The transverse momentum ($p_{\rm T}$) spectra of final state particles are significant observables as they can be utilized to investigate the dynamics. In many studies, searching for a universal scaling of the $p_{\rm T}$ spectra versus suitable variables has been utilized to reveal the dynamics. In ref. \cite{pion_spectrum}, in Au-Au collisions with a center of mass energy per nucleon pair  ($\sqrt{s_{\rm NN}}$) 200 GeV, when presented as a function of $z=p_{\rm T}/K$, the pion spectra exhibited a scaling behaviour independent of the centrality. Here, $K$ was the scaling parameter depending on the centrality. Similar scaling behaviour later was found in the pion spectra at noncentral regions in Au-Au and d-Au collisions, and in the proton and antiproton spectra at midrapidity in Au-Au collisions \cite{non_central_collision,proton_antiproton_spectra}.

In our previous work \cite{inclusive_scaling, pi_k_p_scaling, pp_strange}, the $p_{\rm T}$ spectra of mesons (charged pions and kaons, $K_{S}^{0}$, ${\rm \phi}$) and baryons (protons, ${\rm \Lambda}$, ${\rm \Xi}$) in proton-proton (pp) collisions at the Large Hadron Collider (LHC) have exhibited a scaling behavior independent of the center of mass energy ($\sqrt{s}$) of collisions. It arose when a transformation $p_{\rm T}\rightarrow p_{\rm T}/K$ was applied to the spectra. Here, $K$ was a scaling parameter dependent on $\sqrt{s}$. The scaling behavior of the meson and baryon $p_{\rm T}$ spectra in pp collisions could be explained well in the framework of the colour string percolation (CSP) model \cite{string_perco_model,string_perco_model_0,string_perco_model_1, string_perco_model_2}.

In recent years, universal scaling was observed in the spectra of hadrons produced in nucleus-nucleus (AA)  collisions at the LHC energy scale.  In ref. \cite{AA_scaling_tau}, the authors showed that by defining a suitable saturation momentum $Q_{s}$, the $p_{\rm T}$ spectra in lead-lead (Pb-Pb) collisions with $\sqrt{s_{\rm NN}}=2.76$ TeV at different centralities only depend on $\tau=p_{\rm T}^{2}/Q_{s}^{2}$ in the region with $p_{\rm T}<Q_{s}$. It was argued that the universal scaling was naturally incorporated in the framework of the colour glass condensate mechanism\cite{CGC_1,CGC_2}. In our recent work \cite{Pb_Pb_pi_k_p}, we showed that in the low $p_{\rm T}$ region a scaling behavior independent of the collision centrality was exhibited in the pion, kaon and proton $p_{\rm T}$ spectra in Pb-Pb collisions with $\sqrt{s_{\rm NN}}=$ 2.76 TeV at the LHC.  This scaling behaviour was successfully explained by the CSP model.

Besides pp and AA collisions, there were also proton-nucleus (pA) collisions performed at the LHC. The size of the system produced in pA collisions is intermediate between the sizes of the systems created in pp and AA collisions. Measurement of particle production in pA collisions has frequently been utilized as a baseline to understand the data in AA collisions. In recent years, the mesons (pions and kaons) and baryons (protons, ${\rm \Lambda}$, ${\rm \Xi}$ and ${\rm \Omega}$) $p_{\rm T}$ spectra at different centrality classes (0-5$\%$, 5-10$\%$, 10-20$\%$, 20-40$\%$, 40-60$\%$, 60-80$\%$ and 80-100$\%$) \footnote{Unless otherwise stated, in the later of the paper, we will follow the same definition of centrality classes.} in proton-lead (p-Pb) collisions at $\sqrt{s_{\rm NN}}=$ 5.02 TeV have been published by the ALICE collaboration\cite{data_PI_K_P,data_lambda_Ks0,data_XI_Omega}. As the scaling behaviour independent of the centrality of collisions was observed in the pion, kaon and proton $p_{\rm T}$ spectra in Pb-Pb collisions at $\sqrt{s_{\rm NN}}=$ 2.76 TeV, it is interesting to ask whether a similar scaling behaviour is exhibited in the pion, kaon, proton, ${\rm \Lambda}$, ${\rm \Xi}$ and ${\rm \Omega}$ $p_{\rm T}$ spectra in p-Pb collisions at  5.02 TeV. If the scaling behaviour exists, then we would like to check whether the dependence of the scaling parameter $K$ on the centrality class is similar to that on the centrality in Pb-Pb collisions and the CSP model applied in Pb-Pb collisions can be adopted in the explanation of the scaling behaviour in p-Pb collisions. As the scaling behaviour independent of the center of mass energy is observed in the hadron spectra at pp collisions, we wonder whether there is a similar scaling behaviour independent of the center of mass energy per nucleon pair in p-Pb collisions. However, although the p-Pb collisions with $\sqrt{s_{\rm NN}}=$ 8.16 TeV were performed at the LHC \cite{p_Pb_8_16}, the $p_{\rm T}$ spectra of final state particles at this energy scale are not available so far. Thus, in this work, we do not consider the scaling behaviour of the spectra independent of $\sqrt{s_{\rm NN}}$ in p-Pb collisions.

The organization of this paper is as follows. In sect. \ref{sec:method}, the method to search for the scaling behaviour will be briefly described. In sect. \ref{sec:scaling behaviour}, the scaling behavior in the meson and baryon $p_{\rm T}$ spectra will be shown. In sect. \ref{sec:mechanism}, discussions about the scaling behaviour in the framework of the CSP model will be made. Finally, we will give the conclusion in sect. \ref{sec:conclusion}.

\section{Method to search for the universal scaling}
\label{sec:method}
The method to search for the universal scaling of the meson and baryon spectra in p-Pb collisions at different centrality classes is similar to that in ref. \cite{Pb_Pb_pi_k_p}. When presented in a suitable variable $z=p_{\rm T}/K$, the scaled pion $p_{\rm T}$ spectra at different centrality classes, ${\rm \Phi}(z)=A(\langle N_{\rm{part}}\rangle 2\pi p_{\rm T})^{-1}d^{2}N/dp_{\rm T}dy|_{p_{\rm T}=Kz}$, will exhibit a universal scaling behaviour. Here $\langle N_{\rm{part}}\rangle$ is the average value of the number of participating nucleons,  $(\langle N_{\rm{part}}\rangle 2\pi p_{\rm T})^{-1}d^{2}N/dp_{\rm T}dy$ is the pion's invariant yield per participating nucleon, $y$ is the pion's rapidity. $K$ and $A$ are scaling parameters relying on the centrality class. By shrinking $p_{\rm T}$ with a suitable $K$ and shifting the spectra with another appropriate $A$, the data points at different centrality classes can be put into one curve. Conventionally, we set both $K$ and $A$ as 1 at the most central collisions (the 0-5$\%$ centrality class). With this choice, ${\rm \Phi}(z)$ is exactly the $p_{\rm T}$ spectrum at this centrality class. $K$ and $A$ at other centrality classes will be evaluated with the quality factor (QF) method \cite{QF_1,QF_2}. Apparently, with different choices of $K$ and $A$ at the 0-5$\%$ centrality class, we get different scaling functions ${\rm \Phi}(z)$.  This arbitrariness will disappear if we introduce another scaling variable, $u=z/\langle z\rangle=p_{\rm T}/\langle p_{\rm T} \rangle$, where $\langle z \rangle=\int^{\infty}_{0}z{\rm \Phi}(z)zdz\big/\int^{\infty}_{0}{\rm \Phi}(z)zdz$, and the corresponding normalized scaling function ${\rm \Psi}(u)=\langle z \rangle^{2}{\rm\Phi}(\langle z \rangle u)\big/\int^{\infty}_{0}{\rm \Phi}(z)zdz$. With ${\rm \Psi}(u)$, we can parameterize the $p_{\rm T}$ spectra at other six centrality classes as $f(p_{\rm{T}})=\langle N_{\rm{part}}\rangle/({A}{\langle z \rangle^{2}})\int^{\infty}_{0}{\rm \Phi}(z)zdz \mathrm {\Psi}({p_{\rm{T}}}/({K \langle z \rangle}))$, where $K$ and $A$ are the scaling parameters at these centrality classes. The approaches to search for the universal scaling of the kaon, proton, ${\rm \Lambda}$, ${\rm \Xi}$ and ${\rm \Omega}$ $p_{\rm T}$ spectra are identical to that for the pion spectra.

\section{Universal scaling of the meson and baryon $p_{\rm T}$ spectra}
\label{sec:scaling behaviour}

As shown in sect. \ref{sec:intro}, the ALICE collaboration have published the pion, kaon, proton, ${\rm \Lambda}$, ${\rm \Xi}$ and ${\rm \Omega}$ $p_{\rm T}$ spectra at different centrality classes in p-Pb collisions at $\sqrt{s_{\rm NN}}$ = 5.02 TeV in refs. \cite{data_PI_K_P,data_lambda_Ks0,data_XI_Omega}. Here the pion, kaon, proton, ${\rm \Lambda}$, ${\rm \Xi}$ and ${\rm \Omega}$ $p_{\rm T}$ spectra respectively refer to the spectra of $\pi^{+}+\pi^{-}$, $K^{+}+K^{-}$, $p+\bar{p}$, ${\rm \Lambda}+{\rm \bar \Lambda}$, ${\rm \Xi^{+}}+{\rm \Xi^{-}}$ and ${\rm \Omega^{+}}+{\rm \Omega^{-}}$ per participating nucleon. $\langle N_{\rm{part}}\rangle$ for each centrality class is obtained from ref. \cite{data_Npart}. The pion (kaon, proton, ${\rm \Lambda}$, ${\rm \Xi}$ and ${\rm \Omega}$) spectra at different centrality classes cover a $p_{\rm T}$ range up to 19 (17.5, 17.5, 7, 6.6 and 4.4) GeV/c. Since the scaling parameters $K$ and $A$ are chosen to be 1 at the 0-5$\%$ centrality class, the scaling function ${\rm \Phi}(z)$ is nothing but the $p_{\rm T}$ spectrum at this centrality class.   In ref. \cite{CMS_pi_k_p}, the CMS collaboration have published the pion, kaon and proton spectra covering a  $p_{\rm T}$  range up to 1.2, 1.05 and 1.7 GeV/c respectively in p-Pb collisions at $\sqrt{s_{\rm NN}}$ = 5.02 TeV. These spectra were well fitted by Tsallis distributions \cite{Tsallis_distribution},

\begin{eqnarray}
\label{eq:Tsallis_distribution_p-Pb}
E\frac{d^{3}N}{dp^{3}}= C\left(1-(1-q)\frac{E_{\rm{T}}}{T}\right)^{\frac{1}{1-q}},
\end{eqnarray}
where $C$, $q$ and $T$ are free parameters in the fit, $E_{\rm{T}} =  \sqrt{m^{2}+p_{\rm{T}}^{2}}-m$, $m$ is the particle's mass. The parameter $q$ measures the non-extensivity of the hadronizing system and $1/(q-1)$ determines the power law behaviour of the spectra in the high $p_{\rm T}$ region. The parameter $T$ represents the temperature of the system and controls the exponential behaviour in the low $p_{\rm T}$ region. We find that the formula in eq. (\ref{eq:Tsallis_distribution_p-Pb}) is able to describe the strange particles (${\rm \Lambda}$, ${\rm \Xi}$ and ${\rm \Omega}$)  spectra, but fails to depict the pion, kaon and proton spectra published by the ALICE collaboration in p-Pb collisions at 5.02 TeV. In order to describe the pion, kaon and proton spectra, a double-Tsallis distribution \cite{Pb_Pb_pi_k_p, double_Tsallis_distribution},
\begin{eqnarray}
\label{eq:double_Tsallis_distribution}
E\frac{d^{3}N}{dp^{3}} = C_{1}\left(1-(1-{q_{1}})\frac{E_{\rm{T}}}{T_{1}}\right)^{\frac{1}{1-q_{1}}}+C_{2}\left(1-(1-{q_{2}})\frac{E_{\rm{T}}}{T_{2}}\right)^{\frac{1}{1-q_{2}}},
\end{eqnarray}
is utilized. Here, the first (second) Tsallis distribution represents the soft (hard) yield. Therefore, the scaling function $\mathrm{\Phi}(z)$ for the pion, kaon and proton spectra can be parameterized as
\begin{eqnarray}
\label{eq:phi_z_double_Tsallis_distribution}
{\rm\Phi}(z) &=& C_{1}\left[1-(1-{q_{1}})\frac{\sqrt{m^{2}+z^{2}}-m}{T_{1}}\right]^{\frac{1}{1-q_{1}}}+C_{2}\left[1-(1-{q_{2}})\frac{\sqrt{m^{2}+z^{2}}-m}{T_{2}}\right]^{\frac{1}{1-q_{2}}},
\end{eqnarray}
where $C_{1,2}$, $q_{1,2}$ and $T_{1,2}$ are free parameters. These free parameters are determined by fitting the corresponding $p_{\rm T}$ spectra at the 0-5$\%$ centrality class respectively with eq. (\ref{eq:phi_z_double_Tsallis_distribution}) using the least $\chi^{2}$s method. In the fits the square root of the sum of the statistical and systematic uncertainties of data points has been taken into account. Table  \ref{tab:id_particles_fit_parameters} tabulates these free parameters, their uncertainties and the $\chi^2$s over number of degrees of freedom ($ndf$), named reduced $\chi^{2}$s for the fits.  For protons, the value of $q_{1}$ is very close to be 1. As described in ref. \cite{Tsallis_distribution_1}, when $q_{1}$ tends to be 1, the first Tsallis distribution in eq. (\ref{eq:phi_z_double_Tsallis_distribution}) tends to be an exponential distribution, $C_{1} \mathrm{exp}(-(\sqrt{m^{2}+z^{2}}-m)/T_{1})$. Thus we set $q_{1}$ to be 1 and redo the fit to the proton spectrum at the 0-5$\%$ centrality class. The fit parameters are listed in the fourth column of the upper panel in Table \ref{tab:id_particles_fit_parameters}. From the table, we observe that the non-extensivity parameter $q_{1}$ (the temperature parameter $T_{1}$) in the soft yield of pions, kaons and protons decreases (increases) with the particle's mass. Identical result was found in parameterization of the spectra with the single Tsallis distribution in pp collisions in ref. \cite{pi_k_p_scaling}.

\begin{table}[H]
  \caption{\label{tab:id_particles_fit_parameters} Upper (lower) panel: $C_{1,2}$, $q_{1,2}$ and $T_{1,2}$ ($C_{3}$, $q_{3}$ and $T_{3}$) of ${\rm \Phi}(z)$ for the pion, kaon and proton (${\rm \Lambda}$, ${\rm \Xi}$ and ${\rm \Omega}$) spectra. The errors quoted are due to the quadratic sum of statistical plus systematic uncertainties of data points. In each panel, the last row shows the reduced $\chi^{2}$s.}
\begin{center}
\begin{tabular}{@{}cccc}
\toprule {\ } & \textrm{Pions}   & \textrm{Kaons} & \textrm{Protons} \\
\hline
$C_{1}$ & 6.512$\pm$0.109 & 0.066$\pm$0.018 &0.026$\pm$0.002\\
$q_{1}$ & 1.199$\pm$0.002 & 1.027$\pm$0.019  & 1(fixed)\\
$T_{1}$ & 0.072$\pm$0.003 & 0.397$\pm$0.032 &0.477$\pm$0.006\\
$C_{2}$ & 2.367$\pm$0.095 &0.127$\pm$0.020 & 0.007$\pm$0.003\\
$q_{2}$ & 1.108$\pm$0.001 & 1.139$\pm$0.003 & 1.095$\pm$0.008\\
$T_{2}$ & 0.232$\pm$0.003 & 0.271$\pm$0.006 & 0.449$\pm$0.046\\
$\chi^{2}/ndf$ & 5.76/52 & 12.74/45 & 13.12/44\\
\hline {\ } & \textrm{$\rm \Lambda$}  &\textrm{$\rm \Xi$} &\textrm{$\rm \Omega$} \\
\hline
$C_{3}$ & (199$\pm$8)$\times 10^{-4}$ & (208$\pm$8)$\times10^{-5}$ & (175$\pm$22)$\times10^{-6}$\\
$q_{3}$ & 1.028$\pm$0.008     & 1.002$\pm$0.009  & 1(fixed)\\
$T_{3}$ & 0.501$\pm$0.014     &0.603$\pm$0.018  & 0.663$\pm$0.074\\
$\chi^{2}/ndf$ & 25.59/17 & 17.57/14       & 1.71/5\\
\toprule
\end{tabular}
\end{center}
\end{table}

The scaling function ${\rm \Phi}(z)$ of the ${\rm \Lambda}$, ${\rm \Xi}$ and ${\rm \Omega}$ spectra is parameterized using the single Tsallis formula,
\begin{eqnarray}
 {\rm\Phi}(z)=C_{3}\left[1-(1-q_{3})\frac{\sqrt{m^{2}+z^{2}}-m}{T_{3}}\right]^{\frac{1}{1-q_{3}}},
\label{eq:single_Tsallis_distribution_p_Pb}
\end{eqnarray}
where $C_{3}$, $q_{3}$ and $T_{3}$ are free parameters. These parameters are determined by fitting eq. (\ref{eq:single_Tsallis_distribution_p_Pb}) to the corresponding $p_{\rm T}$ spectra at the 0-5$\%$ centrality class respectively with the least $\chi^{2}$s method. The statistical and systematic uncertainties of data points have been added in quadrature in the fits. In the lower panel of Table \ref{tab:id_particles_fit_parameters}, we present these free parameters, their uncertainties and reduced $\chi^{2}$s for the fits. For $\rm \Omega$, the value of $q_{3}$ tends to be 1. Thus we fix $q_{3}$ as 1 and redo the fit to the $\rm \Omega$ spectrum at the 0-5$\%$ centrality class. The fit parameters are tabulated in the fourth column of the lower panel in Table \ref{tab:id_particles_fit_parameters}. The non-extensivity parameter $q_{3}$ (the temperature parameter $T_{3}$) of the strange baryon spectra also decreases (increases) with the particle's mass.

As described in sect. \ref{sec:method}, we utilize the QF method to evaluate the scaling parameters $K$ and $A$ at other centrality classes. Compared with the method of using the least $\chi^{2}$s fit of the scaled spectra at other centrality classes to the scaling function $\Phi(z)$ in ref. \cite{inclusive_scaling}, this method is more robust since it does not rely on the shape of the scaling function. In this method, a set of data points ($\rho^{i}, \tau^{i}$) is utilized to define the QF \cite{QF_1,QF_2},
\begin{eqnarray}
\textrm{QF}(K,A)=\left[\sum_{i=2}^{n}\frac{(\tau^{i}-\tau^{i-1})^{2}}{(\rho^{i}-\rho^{i-1})^{2}+1/n^{2}}\right]^{-1},
\label{eq:QF_definition}
\end{eqnarray}
where $\rho^{i}=p_{\rm T}^{i}/K$, $\tau^{i}=\textrm{log}(A(2\pi p^{i}_{\rm T})^{-1}d^{2}N^{i}/dp^{i}_{\rm T}dy^{i})$, $n$ is the number of data points. $\rho^{i}$ are ordered and $\tau^{i}$ are rescaled to be in the range between 0 and 1. In order to keep the sum in eq. (\ref{eq:QF_definition}) finite in the case of two successive points having the same $\rho$ values,  a small arbitrary term $1/n^{2} $ is introduced. Obviously, if two successive data points are close in $\rho$ and far in $\tau$, then they will give a large contribution to the sum in eq. (\ref{eq:QF_definition}). Thus, we expect that the data points lie close to a unique curve if they have a small sum (a large QF). The best choice of ($K$, $A$) at other centrality classes is determined to be the one globally maximizing the QF of data points at other centrality classes and the 0-5$\%$ centrality class. Table \ref{tab:a_k_parameters} lists $K$ and $A$ for the pion, kaon, proton, ${\rm \Lambda}$ and ${\rm \Xi}$ spectra at other centrality classes. Also tabulated in the table are the maximum quality factors (QF$_{\rm max}$). For the $\rm \Omega$ spectrum at the 80-100$\%$ centrality class, as there are only four data points, we do not consider it in this work.

\begin{figure}[H]
\centering
\resizebox{0.42\textwidth}{!}{\includegraphics{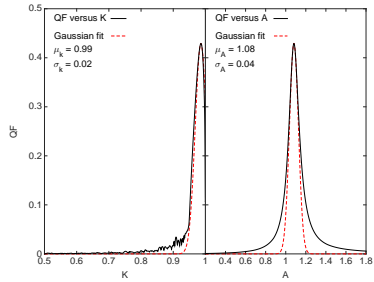}}
\caption{\label{fig:QF_vs_a_k} Left (right) panel: QF as a function of  $K$ ($A$) for the pion spectrum at the 5-10$\%$ centrality class, with $A$ ($K$) fixed to 1.08 (0.99). The black solid (red dash) curve represents the QF scatter plot (the Gaussian fit of the peak with $\rm QF>0.42$).}
\end{figure}

\begin{table}[H]
  \caption{\label{tab:a_k_parameters} $K$ and $A$ for the pion, kaon, proton, ${\rm \Lambda}$ and ${\rm \Xi}$ (${\rm \Omega}$) spectra at the 5-10$\%$, 10-20$\%$, 20-40$\%$, 40-60$\%$, 60-80$\%$ and 80-100$\%$ (5-10$\%$, 10-20$\%$, 20-40$\%$, 40-60$\%$ and 60-80$\%$) centrality classes. The $\rm QF_{max}$ is shown in the last column of this table. The standard deviations of Gaussian fits to the peaks of the QF scatter plots are taken as the uncertainties of $K$ and $A$.}
\begin{center}
\begin{tabular}{@{}ccccc}
\toprule  
\textrm{\ }&
{Centrality}  & 
\textrm{$K$} & 
\textrm{$A$} &
$\rm QF_{max}$\\
\hline
\multirow {6}{*}{\rm Pions}
&5-10$\%$& 0.99$\pm$0.02& 1.08$\pm$0.04&  0.43\\
&10-20$\%$& 0.97$\pm$0.02& 1.12$\pm$0.06& 0.37\\
&20-40$\%$& 0.96$\pm$0.02& 1.14$\pm$0.07& 0.35\\
&40-60$\%$& 0.96$\pm$0.02& 1.16$\pm$0.08& 0.25\\
&60-80$\%$& 0.87$\pm$0.01&1.00$\pm$0.06& 0.24\\
&80-100$\%$& 0.77$\pm$0.01& 1.11$\pm$0.10& 0.14\\
\hline
\multirow {6}{*}{\rm Kaons}
&5-10$\%$& 0.98$\pm$0.01& 1.10$\pm$0.04& 0.10\\
&10-20$\%$& 0.97$\pm$0.01& 1.15$\pm$0.05& 0.84\\
& 20-40$\%$& 0.96$\pm$0.01& 1.20$\pm$0.05& 0.71\\
&40-60$\%$& 0.88$\pm$0.01& 1.04$\pm$0.05& 0.46\\
&60-80$\%$& 0.85$\pm$0.01& 1.07$\pm$0.07&0.31\\
&80-100$\%$&0.75$\pm$0.01& 1.21$\pm$0.08&0.42\\
\hline
\multirow {6}{*}{$\rm Protons$}
&5-10$\%$& 0.98$\pm$0.01& 1.05$\pm$0.04& 1.79\\
&10-20$\%$& 0.96$\pm$0.01& 1.08$\pm$0.04& 1.49\\
& 20-40$\%$& 0.92$\pm$0.01& 1.05$\pm$0.04& 0.80\\
&40-60$\%$& 0.86$\pm$0.01& 0.91$\pm$0.04& 0.74\\
&60-80$\%$& 0.790$\pm$0.005& 0.81$\pm$0.05& 0.43\\
&80-100$\%$& 0.661$\pm$0.004& 0.83$\pm$0.07& 0.32\\
\hline
\multirow {6}{*}{$\rm \Lambda$}
&5-10$\%$& 0.98$\pm$0.02&1.09$\pm$0.06&  2.87\\
&10-20$\%$& 0.96$\pm$0.02& 1.12$\pm$0.09&  2.51\\
& 20-40$\%$& 0.92$\pm$0.03& 1.10$\pm$0.06& 2.74\\
&40-60$\%$&0.88$\pm$0.01& 1.07$\pm$0.07&  1.70\\
&60-80$\%$& 0.81$\pm$0.02& 0.98$\pm$0.08&  1.75\\
&80-100$\%$& 0.70$\pm$0.02& 1.17$\pm$0.10&  1.89\\
\hline
\multirow {6}{*}{$\rm \Xi$}
&5-10$\%$& 0.95$\pm$0.03& 1.05$\pm$0.08&  3.58\\
&10-20$\%$& 0.93$\pm$0.04& 1.11$\pm$0.07& 4.16\\
& 20-40$\%$& 0.92$\pm$0.04& 1.20$\pm$0.07& 4.75\\
&40-60$\%$& 0.87$\pm$0.02& 1.17$\pm$0.08&  3.61\\
&60-80$\%$& 0.77$\pm$0.02& 1.11$\pm$0.08& 3.84\\
&80-100$\%$&0.71$\pm$0.02& 1.67$\pm$0.05&  2.20\\
\hline
&5-10$\%$& 0.96$\pm$0.04& 1.03$\pm$0.12&  8.63\\
&10-20$\%$& 0.97$\pm$0.04& 1.22$\pm$0.11&  11.58\\
\textrm{${\rm \Omega}$}& 20-40$\%$& 0.94$\pm$0.04& 1.37$\pm$0.17& 10.83\\
&40-60$\%$& 0.87$\pm$0.04& 1.31$\pm$0.17& 9.50\\
&60-80$\%$& 0.74$\pm$0.03& 1.10$\pm$0.12&  11.25\\
\bottomrule
\end{tabular}
\end{center}
\end{table}

Using the method mentioned in ref. \cite{QF_1}, we can determine the uncertainties of $K$ and $A$. As an example, we illustrate how to evaluate the uncertainty of $K$($A$) for the pion spectrum at the 5-10$\%$ centrality class. In Fig. \ref{fig:QF_vs_a_k}, the QF as a function of $K$ ($A$) with $A$ ($K$) fixed to the value 1.08 (0.99) returned by the QF method is plotted. The peak value with $\rm QF >(QF_{max}-0.01)$ shows a good scaling and a Gaussian distribution is fitted to this bump. The uncertainty of $K$ ($A$) then is determined to be the standard deviation of the Gaussian fit, $\sigma_{K(A)}$. As the mean value of the Gaussian fit, $\mu_{K(A)}$, is in agreement with the value of $K$ ($A$) returned by the QF method,  the method to determine the uncertainties of $K$ ($A$) is robust. The uncertainties of $K$ and $A$ for the pion, kaon, proton, ${\rm \Lambda}$, ${\rm \Xi}$ and ${\rm \Omega}$ spectra at other centrality classes  are evaluated by making Gaussian fits to their QF peaks with $\rm QF>(QF_{max}-0.01)$.

\begin{figure}[H]
  \centering
\resizebox{0.42\textwidth}{!}{\includegraphics{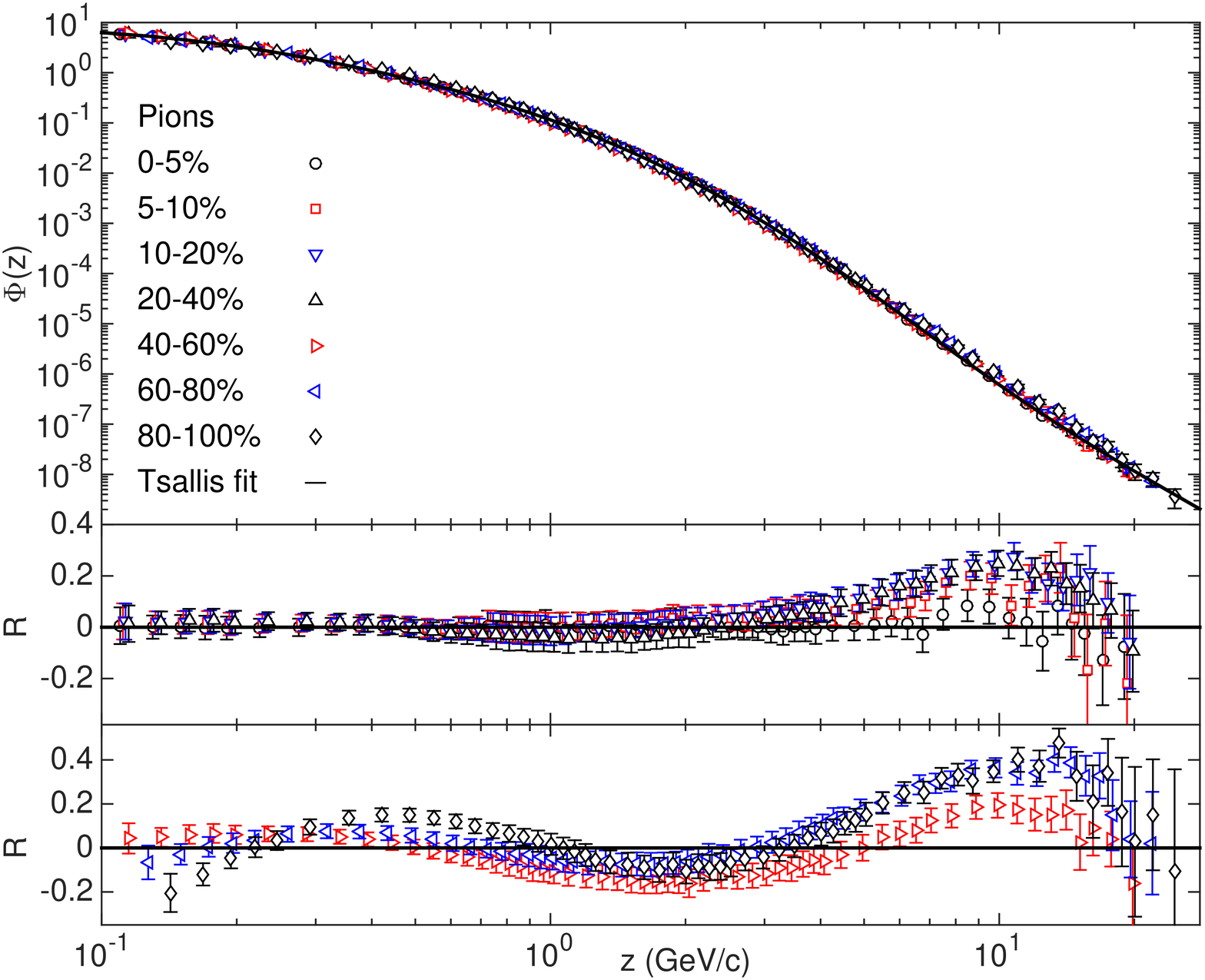}}\put(-36,102){(a)}
\resizebox{0.42\textwidth}{!}{\includegraphics{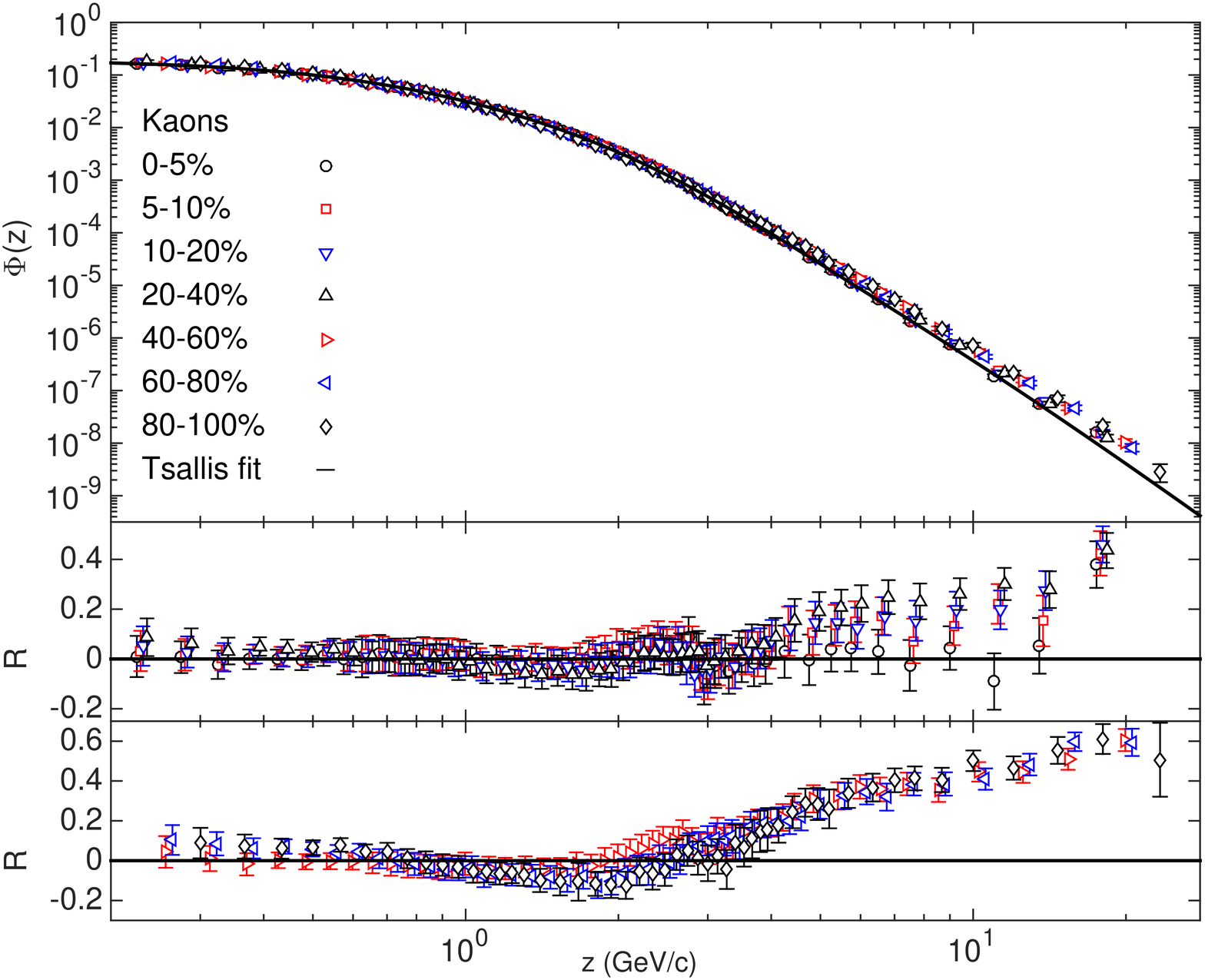}}\put(-36,102){(b)}
\caption{\label{fig:PI_K_PT_spectra} Upper panel in (a) ((b)): the universal scaling of the pion (kaon) $p_{\rm T}$ spectra presented in $z$. The black curve is described by eq. (\ref{eq:phi_z_double_Tsallis_distribution}) with parameters in the second (third) column of the upper panel in Table \ref{tab:id_particles_fit_parameters}. The data points are taken from ref. \cite{data_PI_K_P}. The middle (lower) panels in (a) and (b): the $R$ distributions for the spectra at the 0-5$\%$, 5-10$\%$, 10-20$\%$ and 20-40$\%$ (40-60$\%$, 60-80$\%$ and 80-100$\%$) centrality classes.}
\end{figure}

Using $K$ and $A$ in Table \ref{tab:a_k_parameters}, we plot the scaled pion $p_{\rm T}$ spectra at different centrality classes in the upper panel of Fig. \ref{fig:PI_K_PT_spectra} (a). On a logarithmic scale, most of the data points at different centrality classes appear to be shifted to a universal curve within errors. The universal curve is described by the scaling function ${\rm \Phi}(z)$ in eq. \ref{eq:phi_z_double_Tsallis_distribution} with parameters in the second column of the upper panel in Table \ref{tab:id_particles_fit_parameters}. In order to see how the experimental data points agree with the universal curve, a ratio $R=\rm (data-fitted)/data$ is evaluated for the spectra at different centrality classes, as shown in the middle (lower) panel of this figure. For the spectra at the 0-5$\%$ and 40-60$\%$ centrality classes, all the data points have absolute values of $R$ less than 0.2.  For the spectra at the 5-10$\%$, 10-20$\%$, 20-40$\%$ and 60-80$\%$ centrality classes, the $R$ values of the data points in the regions with $z\leq$ 7.6, 6.9, 7.1 and 4.9 ($p_{\rm T}\leq$ 7.5, 6.8, 6.8 and 4.3) GeV/c are in the range from $-0.2$ to 0.2. For the spectrum at the 80-100$\%$ centrality class, except for the first data point, all the other data points in the region with $z\leq$ 5.1 ($p_{\rm T}\leq$ 3.9) GeV/c are consistent with the fitted curve within 20$\%$. Thus, in the region with $z \leq 4.9$ ($p_{\rm{T}} \leq 3.9$) GeV/c, almost all the data points at all centrality classes agree with the fitted curve within 20$\%$.

In the upper panel of Fig. \ref{fig:PI_K_PT_spectra} (b), we present the universal scaling of the kaon spectra. For the spectrum at the 0-5$\%$ centrality class, except for the last data point, all the other data points have $R$ values from -0.2 and 0.2. For the spectra at other centrality classes, the data points with $z\leq$ 9.2, 11.3, 5, 4, 4.1 and 4.1 ($p_{\rm T}\leq$ 9, 11, 4.8, 3.5, 3.5 and 3.1) GeV/c, respectively, agree with the fitted curve within 20$\%$. Therefore, in the region with $z \leq 4$ ($p_{\rm{T}} \leq 3.1$) GeV/c, all the data points at all centrality classes are consistent with the fitted curve within 20$\%$.

\begin{figure}[H]
  \centering
\resizebox{0.42\textwidth}{!}{\includegraphics{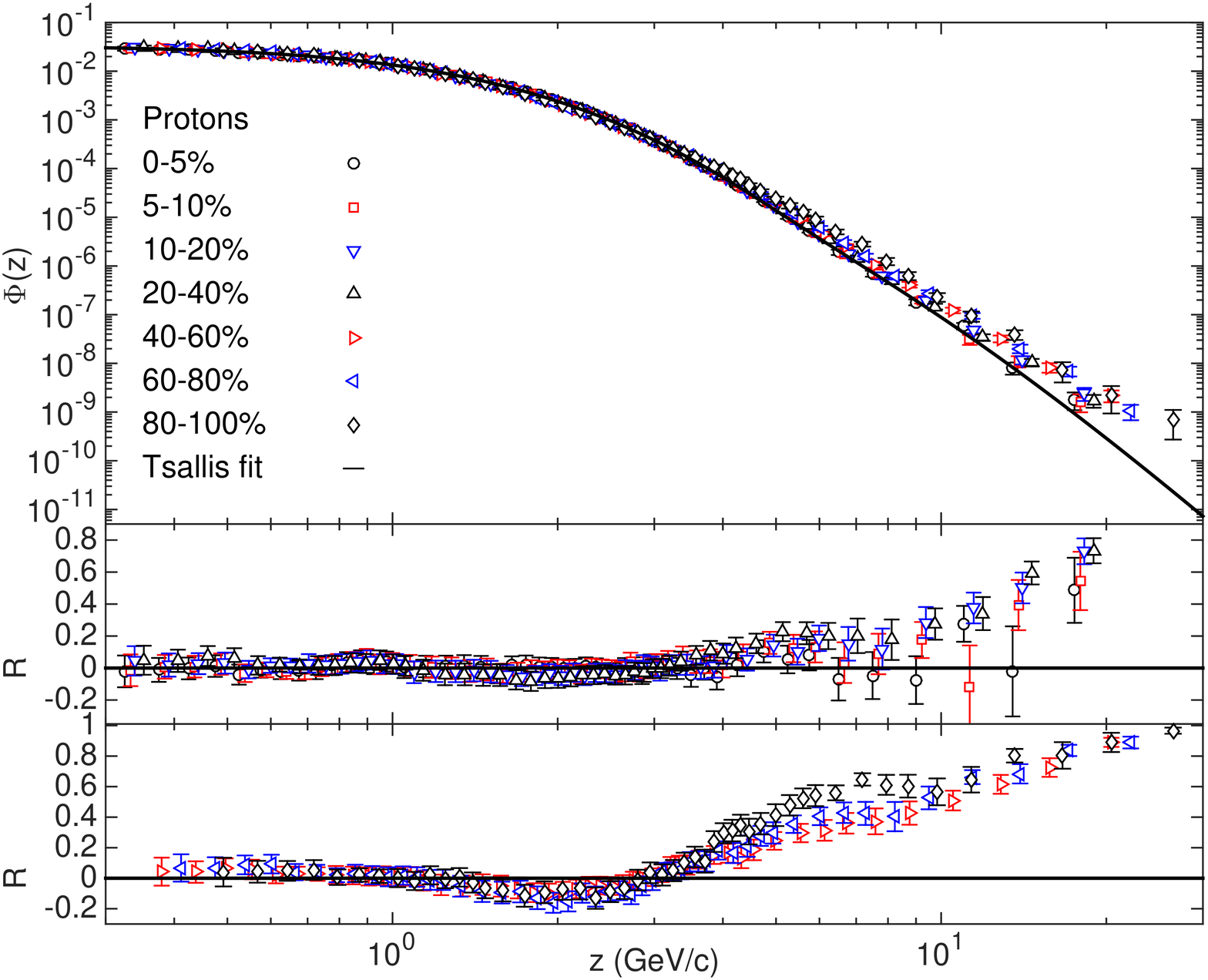}}\put(-36,102){(a)}
\resizebox{0.42\textwidth}{!}{\includegraphics{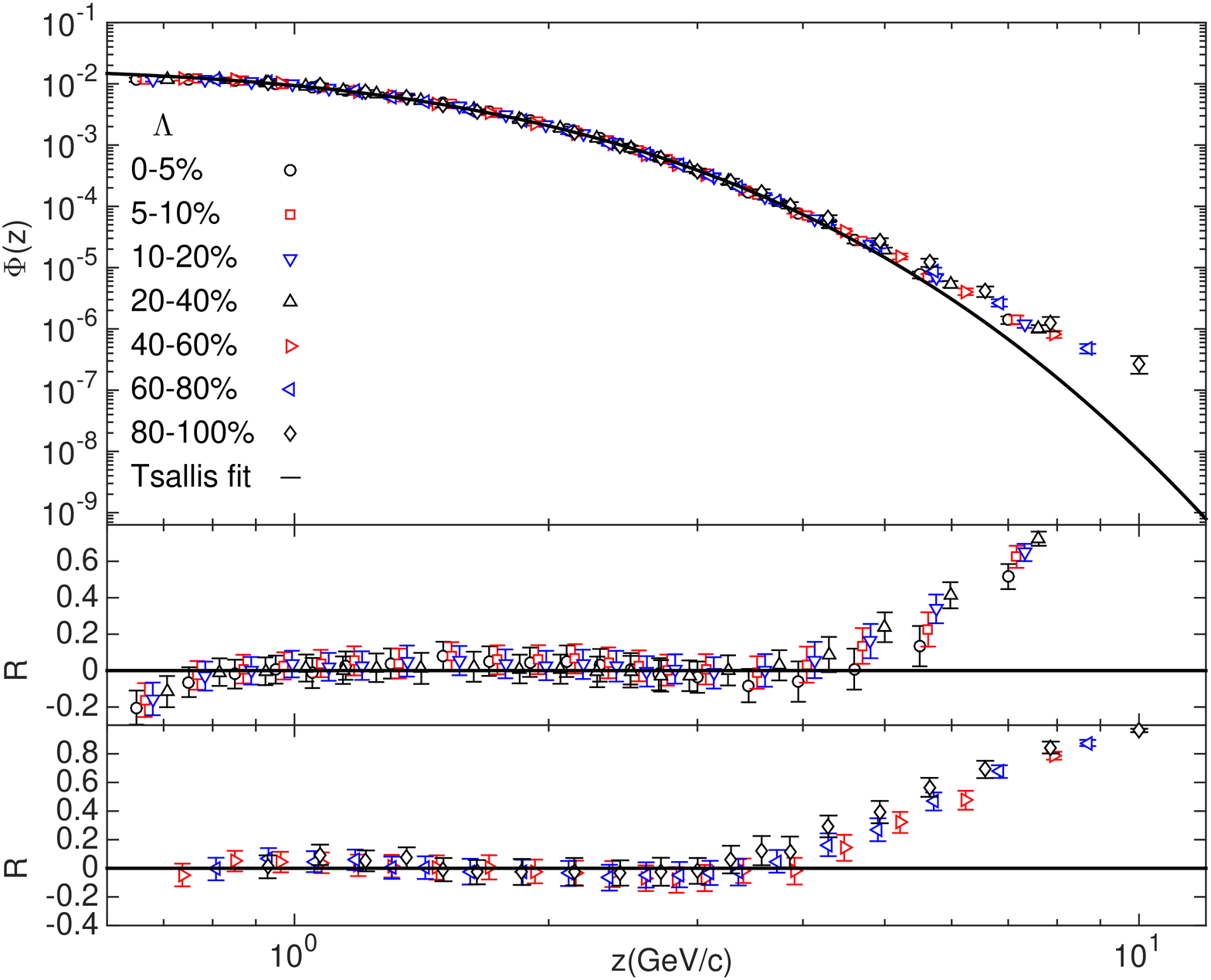}}\put(-36,102){(b)}
\caption{\label{fig:P_Lambda_PT_spectra} Upper panel in (a) ((b)): the universal scaling of the proton ($\mathrm{\Lambda}$) spectra presented in $z$. The black curve is described by eq. (\ref{tab:id_particles_fit_parameters}) ((\ref{eq:single_Tsallis_distribution_p_Pb})) with parameters in the fourth (second) column of the upper (lower) panel in Table \ref{tab:id_particles_fit_parameters}. The data points are taken from refs. \cite{data_PI_K_P, data_lambda_Ks0}. The middle (lower) panels in (a) and (b): the $R$ distributions for the spectra at the 0-5$\%$, 5-10$\%$, 10-20$\%$ and 20-40$\%$ (40-60$\%$, 60-80$\%$ and 80-100$\%$) centrality classes.}
\end{figure}

The universal scaling of the proton spectra is showed in the upper panel of Fig. \ref{fig:P_Lambda_PT_spectra} (a). For the spectra at different centrality classes, the data points with $z\leq$ 9, 11.3, 7.8, 4.6, 4.6, 4.2 and 3.7 ($p_{\rm T}\leq$ 9, 11, 7.5, 4.3, 3.9, 3.3 and 2.5) GeV/c, respectively, have absolute $R$ values smaller than 0.2. As a result, in the region with $z \leq 3.7$ ($p_{\rm{T}} \leq 2.5$) GeV/c, all the data points at all centrality classes are in agreement with the fitted curve within 20$\%$.

In the upper panel of Fig. \ref{fig:P_Lambda_PT_spectra} (b), the universal scaling of the ${\rm \Lambda}$ spectra is presented. For the spectrum at the 0-5$\%$ centrality class, except for the first data point, all the other data points with $z\leq$ 5.5 ($p_{\rm T}\leq$ 5.5) GeV/c have $R$ values in the range from $-$0.2 and 0.2. For the spectra at other centrality classes, the data points with $z\leq$ 4.7, 4.8, 4.3, 4.5, 4.3 and 3.9 ($p_{\rm T}\leq$ 4.6, 4.6, 4, 4, 3.5 and 2.7) GeV/c, respectively,  agree with the fitted curve within 20$\%$. Thus, in the region with $z \leq 3.9$ ($p_{\rm{T}} \leq 2.7$) GeV/c, almost all the data points at all centrality classes are consistent with the fitted curve within 20$\%$.

\begin{figure}[H]
   \centering
\resizebox{0.42\textwidth}{!}{\includegraphics{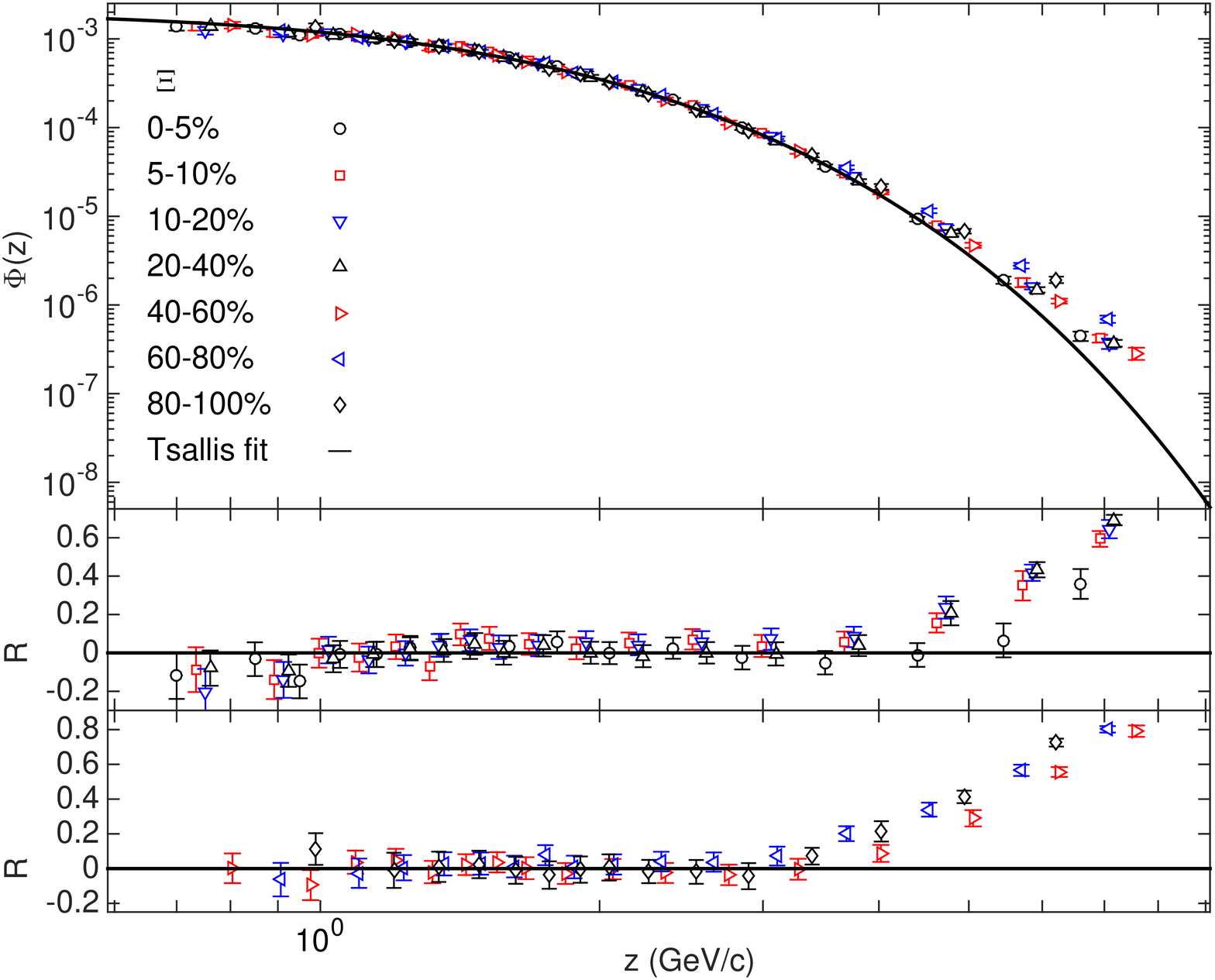}}\put(-36,102){(a)}
\resizebox{0.42\textwidth}{!}{\includegraphics{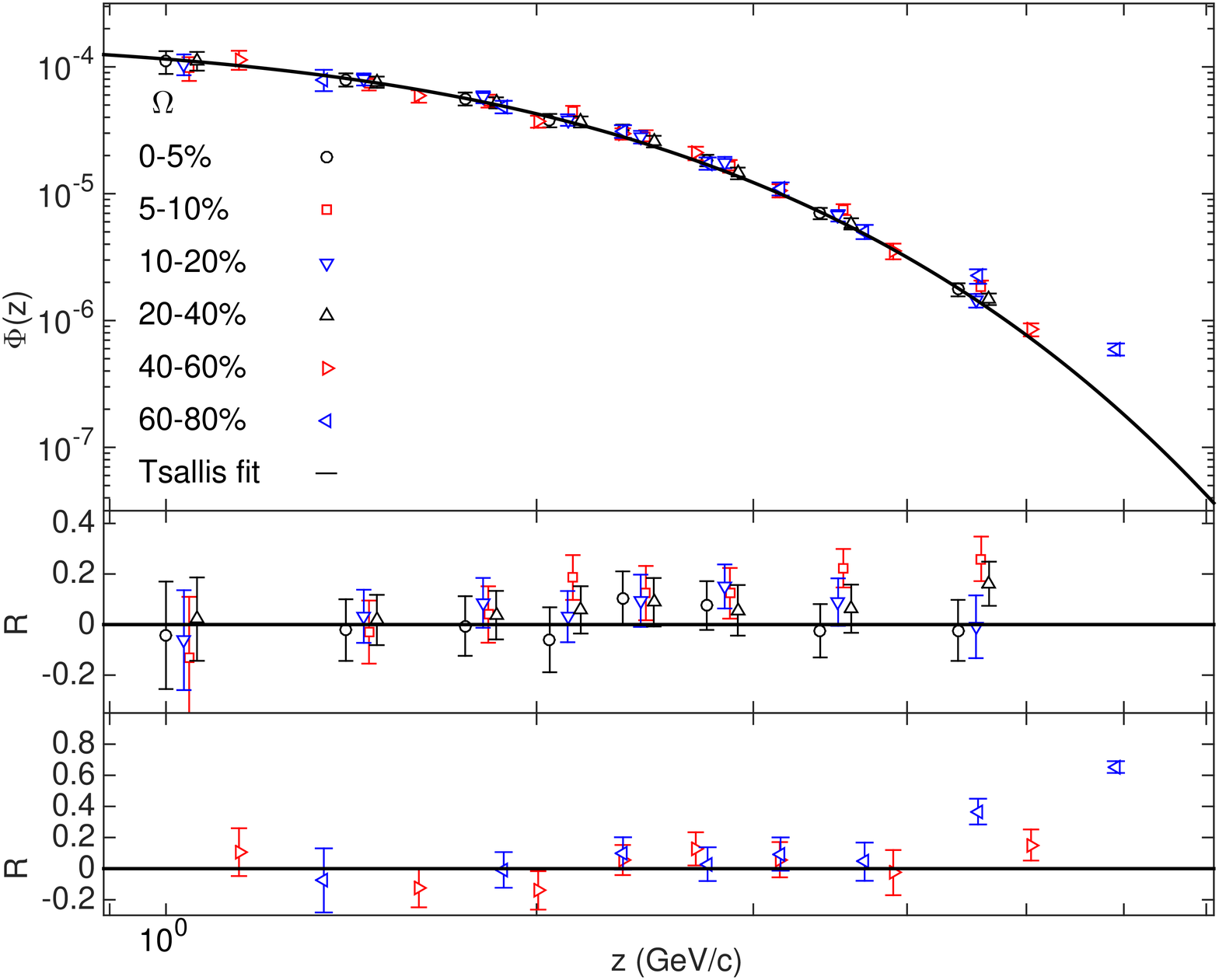}}\put(-36,102){(b)}
\caption{\label{fig:XI_Omega_PT_spectra} Upper panel in (a) ((b)): the universal scaling of the ${\rm \Xi}$ (${\rm \Omega}$) spectra presented in $z$. The black curve is described by eq. (\ref{eq:single_Tsallis_distribution_p_Pb}) with parameters in the third (fourth) column of the lower panel of Table \ref{tab:id_particles_fit_parameters}. The data points are taken from ref.  \cite{data_XI_Omega}. The middle (lower) panels in (a) and (b): the $R$ distributions at the 0-5$\%$, 5-10$\%$, 10-20$\%$ and 20-40$\%$ (40-60$\%$, 60-80$\%$ and 80-100$\%$ (40-60$\%$ and 60-80$\%$) for the ${\rm \Xi}$ (${\rm \Omega}$) spectra) centrality classes.}
\end{figure}

The universal scaling of the ${\rm \Xi}$ spectra is exhibited in the upper panel of Fig. \ref{fig:XI_Omega_PT_spectra} (a). For the spectra at the 0-5$\%$, 5-10$\%$, 20-40$\%$, 40-60$\%$, 60-80$\%$ and 80-100$\%$ centrality classes, the data points with $z\leq$ 5.5, 4.6, 3.8, 4, 3.1 and 3.4 ($p_{\rm T}\leq$ 5.5, 4.4, 3.5, 3.5, 2.4 and 2.4) GeV/c have absolute $R$ values less than 0.2. For the spectrum at the 10-20$\%$ centrality class, except for the first data point, the $R$ values of the data points with $z \leq 3.8$ ($p_{\rm{T}} \leq 3.5$) GeV/c are in the range from $-$0.2 to 0.2. Therefore, in the region with $z \leq 3.1$ ($p_{\rm{T}} \leq 2.4$) GeV/c, almost all the data points at all centrality classes are in agreement with the fitted curve within 20$\%$.

In the upper panel of Fig. \ref{fig:XI_Omega_PT_spectra} (b), we show the universal scaling of the ${\rm \Omega}$ spectra. For the spectrum at the 0-5$\%$ centrality class,  all the data points agree with the fitted curve within 20$\%$. For the spectra at other centrality classes, the data points with $z\leq$ 2.9, 4.6, 4.7, 5 and 3.7 ($p_{\rm T}\leq$ 2.8, 4.4, 4.4, 4.4 and 2.8) GeV/c, respectively, are consistent with the fitted curve within 20$\%$. Thus, in the region with $z \leq 2.9$ ($p_{\rm{T}} \leq 2.8$) GeV/c, all the data points at all centrality classes coincide with the fitted curve within 20$\%$.

As a summary, in the low $p_{\rm T}$ regions with $z \leq 4.9$, 4, 3.7, 3.9, 3.1 and 2.9 ($p_{\rm{T}}\leq 3.9$, 3.1, 2.5, 2.7, 2.4 and 2.8) GeV/, a universal scaling independent of the centrality class is indeed exhibited in the pion, kaon, proton, ${\rm \Lambda}$, ${\rm \Xi}$ and ${\rm \Omega}$ spectra respectively. In these regions, almost all the data points are consistent with the fitted curves within 20$\%$. However, outside these regions, a departure of the proposed scaling is observed going from central to peripheral collisions. The more peripheral the collisions are, the more obvious the violation of the scaling is. This hierarchy of the scaling violation is similar to that observed in Pb-Pb collisions in ref. \cite{Pb_Pb_pi_k_p}. As described in ref. \cite{data_lambda_Ks0}, it may no longer be valid to assume that the final state effects can be ignored in pA collisions at the LHC, since the pseudo-rapidity density of final state particles in pA collisions are comparable to that in semi-peripheral Au-Au and Cu-Cu collisions at top RHIC energy. Moreover,  the theoretical models which incorporate final state effects give a better description of the data in pA collisions \cite{data_lambda_Ks0}. Thus, the violation of the universal scaling in p-Pb collisions may be due to the final state effects.

\begin{figure}[H]
  \centering
\resizebox{0.45\textwidth}{!}{\includegraphics{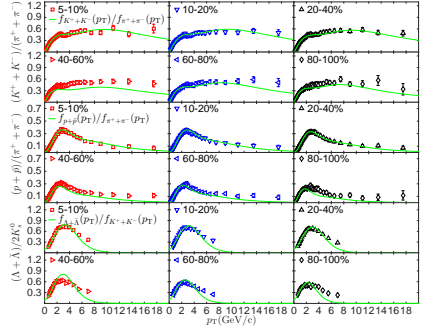}}
\caption{\label{fig:particle_ratios} The panels in the top (middle and bottom) two rows in the figure: $(K^{+}+K^{-})/(\pi^{+}+\pi^{-})$ ($(p+\bar{p})/(\pi^{+}+\pi^{-})$ and $({\rm \Lambda}+\bar{{\rm \Lambda}})/(2{ K}_{S}^{0})$) distributions at the 5-10$\%$, 10-20$\%$, 20-40$\%$, 40-60$\%$, 60-80$\%$ and 80-100$\%$ centrality classes. Green curves represent $f_{K^{+}+K^{-}}(p_{\rm T})/f_{\pi^{+}+\pi^{-}}(p_{\rm T})$ ($f_{p+\bar{p}}(p_{\rm T})/f_{\pi^{+}+\pi^{-}}(p_{\rm T})$ and $f_{{\rm \Lambda}+\bar{{\rm \Lambda}}}(p_{\rm T})/f_{K^{+}+K^{-}}(p_{\rm T})$). The data points are taken from refs. \cite{data_PI_K_P,data_lambda_Ks0}.}
\end{figure}

As described in sect. \ref{sec:method}, the scaling functions ${\rm \Phi}(z)$ in eqs. (\ref{eq:phi_z_double_Tsallis_distribution}) and (\ref{eq:single_Tsallis_distribution_p_Pb}) depend on the choice of $K$ and $A$ at the 0-5$\%$ centrality class. This dependence will be eliminated if the spectra are presented in another variable $u=z/\langle z \rangle$. The values of $\langle z \rangle$ for the pion, kaon, proton, ${\rm \Lambda}$, ${\rm \Xi}$ and ${\rm \Omega}$ spectra are determined to be 0.548$\pm$0.008, 0.947$\pm$0.039, 1.233$\pm$0.041, 1.352$\pm$0.037, 1.546$\pm$0.043 and 1.756$\pm$0.163 GeV/c, where the errors originate from the uncertainties of  $C_{1,2,3}$, $q_{1,2,3}$ and $T_{1,2,3}$ in Table \ref{tab:id_particles_fit_parameters}. By substituting $z$ and ${\rm \Phi}(z)$ into ${\rm \Psi}(u)$,  we can write the normalized scaling functions ${\rm \Psi}(u)$ of the pion, kaon and proton spectra as
\begin{eqnarray}
\label{ normalized}
{\rm\Psi}(u)&=&C_{1}^{'}\left[1-(1-q_{1})\frac{\sqrt{{m^{'}}^{2}+u^{2}}-m^{'}}{u_{1}}\right]^{\frac{1}{1-q_{1}}}+C_{2}^{'}\left[1-(1-{q_{2}})\frac{\sqrt{{m^{'}}^{2}+u^{2}}-m^{'}}{u_{2}}\right]^{\frac{1}{1-q_{2}}}.
\end{eqnarray}
For the ${\rm \Lambda}$, ${\rm \Xi}$ and ${\rm \Omega}$ spectra, the ${\rm \Psi}(u)$ can be written as
\begin{eqnarray}
{\rm \Psi}(u)=C^{\prime}_{3}\left[1-(1-q_{3})\frac{\sqrt{{m^{'}}^{2}+u^{2}}-m^{\prime}}{u_{3}}\right]^{\frac{1}{1-q_{3}}},
\label{eq:psi_u_pt_spectrum_pp}
\end{eqnarray}
where $C^{'}_{1,2,3}=\langle z\rangle^{2}C_{1,2,3}/\int^{\infty}_{0}{\rm \Phi}(z)zdz$,  $u_{1,2,3}=T_{1,2,3}/\langle z\rangle$ and $m^{\prime}=m/\langle z \rangle$. Their values are presented in Table \ref{tab:id_particles_normalized_parameters}. With the normalized scaling function $\mathrm{\Psi}(u)$, we can parameterize the $p_{\rm T}$ spectra at other centrality classes as $f(p_{\rm{T}})=\langle N_{\rm{part}}\rangle/({A}{\langle z \rangle^{2}})\int^{\infty}_{0}{\rm \Phi}(z)zdz\mathrm{\Psi}({p_{\rm{T}}}/({K \langle z \rangle}))$, where $K$ and $A$ are the scaling parameters at these centrality classes. In refs. \cite{data_PI_K_P,data_lambda_Ks0}, the ALICE collaboration have presented the kaon to pion ratio ($(K^{+}+K^{-})/(\pi^{+}+\pi^{-})$), the proton to pion ratio ($(p+\bar{p})/(\pi^{+}+\pi^{-})$) and the $\rm \Lambda$ to ${K}_{S}^{0}$ ratio ($({\rm \Lambda}+\bar{{\rm \Lambda}})/(2{K}_{S}^{0})$) as a function of $p_{\rm T}$ at different centrality classes. In Fig. \ref{fig:particle_ratios}, we show in the low $p_{\rm T}$ region the $(K^{+}+K^{-})/(\pi^{+}+\pi^{-})$ ($(p+\bar{p})/(\pi^{+}+\pi^{-})$ and $({\rm \Lambda}+\bar{{\rm \Lambda}})/(2{ K}_{S}^{0})$) distributions at other centrality classes are well described by $f_{K^{+}+K^{-}}(p_{\rm T})/f_{\pi^{+}+\pi^{-}}(p_{\rm T})$ ($f_{p+\bar{p}}(p_{\rm T})/f_{\pi^{+}+\pi^{-}}(p_{\rm T})$ and $f_{{\rm \Lambda}+\bar{{\rm \Lambda}}}(p_{\rm T})/f_{K^{+}+K^{-}}(p_{\rm T})$)\footnote{As the $K_{S}^{0}$ spectrum is theoretically the same as the $K^{\pm}$ spectrum, the universal scaling of the $K_{S}^{0}$ spectra is similar to that of the $K^{\pm}$ spectra. Thus we use $f_{{\rm \Lambda}+\bar{{\rm \Lambda}}}(p_{\rm T})/f_{K^{+}+K^{-}}(p_{\rm T})$ to replace $f_{{\rm \Lambda}+\bar{{\rm \Lambda}}}(p_{\rm T})/2f_{K_{S}^{0}}(p_{\rm T})$. }. The agreement definitely indicates the existence of the universal scaling  in the low $p_{\rm T}$ region. In the high $p_{\rm T}$ region, an obvious deviation of $f_{K^{+}+K^{-}}(p_{\rm T})/f_{\pi^{+}+\pi^{-}}(p_{\rm T})$ ($f_{p+\bar{p}}(p_{\rm T})/f_{\pi^{+}+\pi^{-}}(p_{\rm T})$ and $f_{{\rm \Lambda}+\bar{{\rm \Lambda}}}(p_{\rm T})/f_{K^{+}+K^{-}}(p_{\rm T})$) and the data points  is observed going from central to peripheral collisions due to the violation of the scaling.

\begin{table}[H]
  \caption{\label{tab:id_particles_normalized_parameters}  $C^{'}_{1,2,3}$, $u_{1,2,3}$ and $m^{'}$ of ${\rm \Psi}(u)$ for the pion, kaon, proton, ${\rm \Lambda}$, ${\rm \Xi}$ and ${\rm \Omega}$ $p_{\rm T}$ spectra. The errors quoted are due to the uncertainties of $C_{1,2,3}$, $q_{1,2,3}$ and $T_{1,2,3}$ in Table \ref{tab:id_particles_fit_parameters}.}
\begin{center}
\begin{tabular}{@{}cccc}
\toprule \textrm{\ } & \textrm{Pions}&  \textrm{Kaons}& \textrm{Protons} \\
\hline
$C^{'}_{1}$  &4.671$\pm$0.135    & 0.994$\pm$0.227    & 1.735$\pm$0.153\\
$u_{1}$      & 0.131$\pm$0.005   &0.419$\pm$0.022     & 0.386$\pm$0.013\\
$C^{'}_{2}$  & 1.698$\pm$ 0.066  & 1.924$\pm$0.299   & 0.475$\pm$0.189\\
$u_{2}$      &0.423$\pm$0.003    & 0.286$\pm$0.012   & 0.364$\pm$0.027\\
$m^{'}$      & 0.255$\pm$0.004   &0.522$\pm$0.021    & 0.761$\pm$0.025\\
\hline
\textrm{\ } & \textrm{$\rm \Lambda$} & \textrm{$\rm \Xi$} & \textrm{$\rm \Omega$}\\
\hline
$C^{'}_{3}$   & 2.15$\pm$0.04    & 2.053$\pm$0.037  & 1.99$\pm$0.06\\
$u_{3}$   & 0.371$\pm$0.006  & 0.390$\pm$0.007   & 0.378$\pm$0.007\\
$m^{'}$   & 0.825$\pm$0.023  & 0.855$\pm$0.024   &0.952$\pm$0.089\\
\toprule
\end{tabular}
\end{center}
\end{table}

\section{Discussions}\label{sec:mechanism}
In this section, we would like to discuss the universal scaling of the meson and baryon spectra in the low $p_{\rm T}$ region with the framework of the CSP model \cite{string_perco_model,string_perco_model_0,string_perco_model_1,string_perco_model_2}. With the framework of this model, in p-Pb collisions colour strings stretched between the partons of the projectile and target will decay into new ones due to the creation of $q\bar{q}$ pairs from vacuum. Subsequent hadronization of these new strings will generate mesons (pions and kaons) and baryons (protons, ${\rm \Lambda}$, ${\rm \Xi}$ and ${\rm \Omega}$). In the transverse plane, the colour in these strings are confined within a finite area of $S_{1}=\pi r_{0}^{2}$, where $r_{0}\approx 0.2$ fm. As described in ref. \cite{string_perco_model,string_perco_model_0,string_perco_model_1}, the number of strings grows as the colliding energy or atomic number of the colliding particles increases. Since these strings carry colour, they interact with each other and overlap to form clusters. A cluster with $n$ ($n>1$) strings is assumed to behave as a single string with a higher colour field $\vv{Q}_{n}$, which is the vectorial sum of the colour charge of each individual $\vv{Q}_{1}$ string. Since the orientations of individual string colour fields are arbitrary, the average value of $\vv{Q}_{1i}\cdot\vv{Q}_{1j}$ is zero and $\vv{Q}_{n}^{2}=n\vv{Q}_{1}^{2}$. As the strings may partially overlap, $\vv{Q}_{n}$ also relies on the transverse areas of the cluster $S_{n}$, $Q_{n}=\sqrt{nS_{n}/S_{1}} Q_{1}$. By the influence of a stronger field in a fused string (the cluster), the average $p_{\rm T}^{2}$, $\langle p_{\rm{T}}^{2}\rangle_{n}$, of particles produced by the cluster is given by $\langle p_{\rm{T}}^{2}\rangle_{n}=\sqrt{nS_{1}/S_{n}} \langle p_{\rm{T}}^{2}\rangle_{1}$, where $\langle p_{\rm{T}}^{2}\rangle_{1}$ is the mean $p_{\rm{T}}^{2}$ of particles produced by a single string, $nS_{1}/S_{n}$ is the degree of string overlap. If strings just touches with each other, then $S_{n}=nS_{1}$, $nS_{1}/S_{n}=1$ and $\langle p_{\rm T}^{2}\rangle_{n}=\langle p_{\rm T}^{2}\rangle_{1}$, which means that the $n$ strings decay into hadrons independently. If strings fully overlap with each other, then $S_{n}=S_{1}$, $nS_{1}/S_{n}=n$ and $\langle p_{\rm T}^{2}\rangle_{n}=\sqrt{n} \langle p_{\rm T}^{2}\rangle_{1}$, which means that the mean $p_{\rm T}^{2}$ is maximally enhanced due to the percolation.

The $p_{\rm{T}}$ spectra of hadrons produced in p-Pb collisions can be written as a superposition of the $p_{\rm{T}}$ spectra produced by each cluster, $g(x, p_{\rm{T}})$,  weighted with the distribution of the size for the clusters, $W(x)$,
\begin{eqnarray}
\frac{d^{2}N}{2\pi p_{\rm{T}}dp_{\rm{T}}dy}=C\int_{0}^{1/p^{2}_{\rm{T}}} W(x) g(x, p_{\rm{T}})dx,
\label{eq:CPS_formula}
\end{eqnarray}
where $C$ is a normalization constant characterizing the total number of clusters formed for hadrons before hadronization. The cluster's size distribution, $W(x)$, is assumed to be a gamma distribution,
\begin{eqnarray}
W(x)=\frac{\gamma}{\Gamma(\kappa)}(\gamma x)^{\kappa-1}\textrm{exp}(-\gamma x),
\label{eq:gamma_function}
\end{eqnarray}
where $x$ is proportional to $1/\langle p_{\rm T}^{2}\rangle_{n}$, $\kappa$ and $\gamma$ are free parameters. $\kappa$ is proportional to the inverse of the dispersion of the size distribution, $1/\kappa=(\langle x^{2}\rangle-\langle x\rangle^{2})/\langle x\rangle^{2}$. $\gamma$ is related to the mean value of $x$, $\langle x\rangle = \kappa/\gamma$. When the CSP model was proposed in refs. \cite{string_perco_model_1,string_perco_model_2}, the cluster's fragmentation function was originally assumed to be the Schwinger formula $g(x, p_{\rm{T}})=\mathrm{exp}(-p_{\rm T}^{2} x)$. However, with this fragmentation function, eq. (\ref{eq:CPS_formula}) fails to describe the spectra in p-Pb collisions. Thus as done in ref. \cite{frag_function_variable}, the fragmentation function of the cluster $g(x, p_{\rm{T}})$ is chosen to be analogous to the usual fragmentation functions from hard  partons to hadrons \cite{frag_function_form}
\begin{eqnarray}
 g(x, p_{\rm{T}})=D\xi^{\alpha}(1-\xi)^{\eta}(1+\xi)^{\theta},
\label{eq:frag_function_form_1}
\end{eqnarray}
where $\xi = p_{\rm{T}}\sqrt{x}$ is the fraction of the hadron's transverse momentum relative to that of a cluster with size $x$, $D$, $\alpha$, $\eta$ and $\theta$ are free parameters. The parameter $D$ in eq. (\ref{eq:frag_function_form_1}) can be absorbed into $C$ in eq. (\ref{eq:CPS_formula}), since they always appear as a product. As the maximum fraction of the transverse momentum carried by the hadron is 1, the upper limit of the integration in eq. (\ref{eq:CPS_formula}) is set to be $1/p_{\rm T}^{2}$, rather than infinity. For different hadrons, as they are generated from different decay channels of the clusters, $g(x, p_{\rm{T}})$ should be different. However, for the same hadrons produced from clusters with different sizes $x$, $g(x, p_{\rm{T}})$ should be the same, which means that $g(x, p_{\rm{T}})$ should be universal for all centrality classes in the collisions.

\begin{figure}[H]
   \centering
\resizebox{0.42\textwidth}{!}{\includegraphics{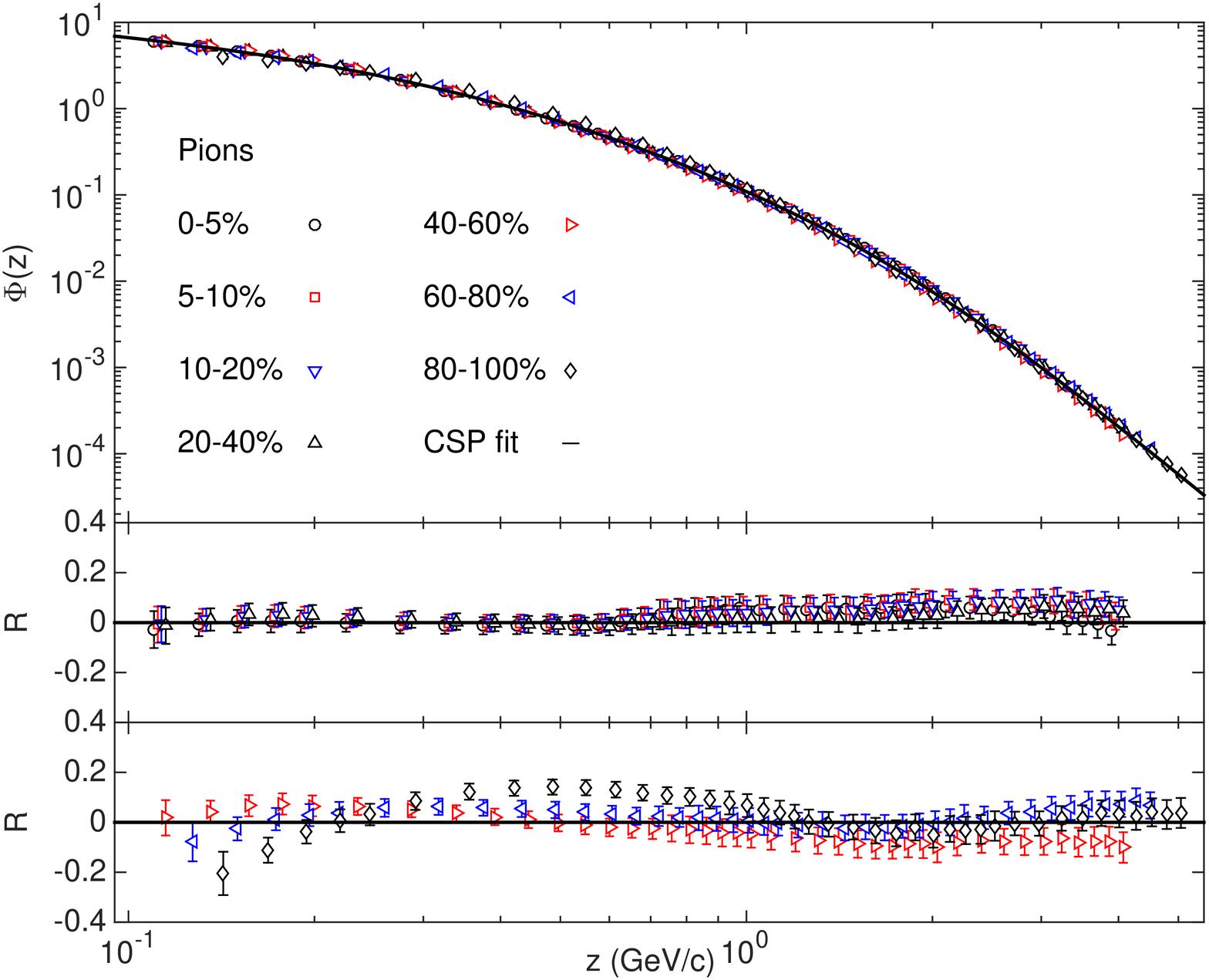}}\put(-36,102){(a)}
\resizebox{0.42\textwidth}{!}{\includegraphics{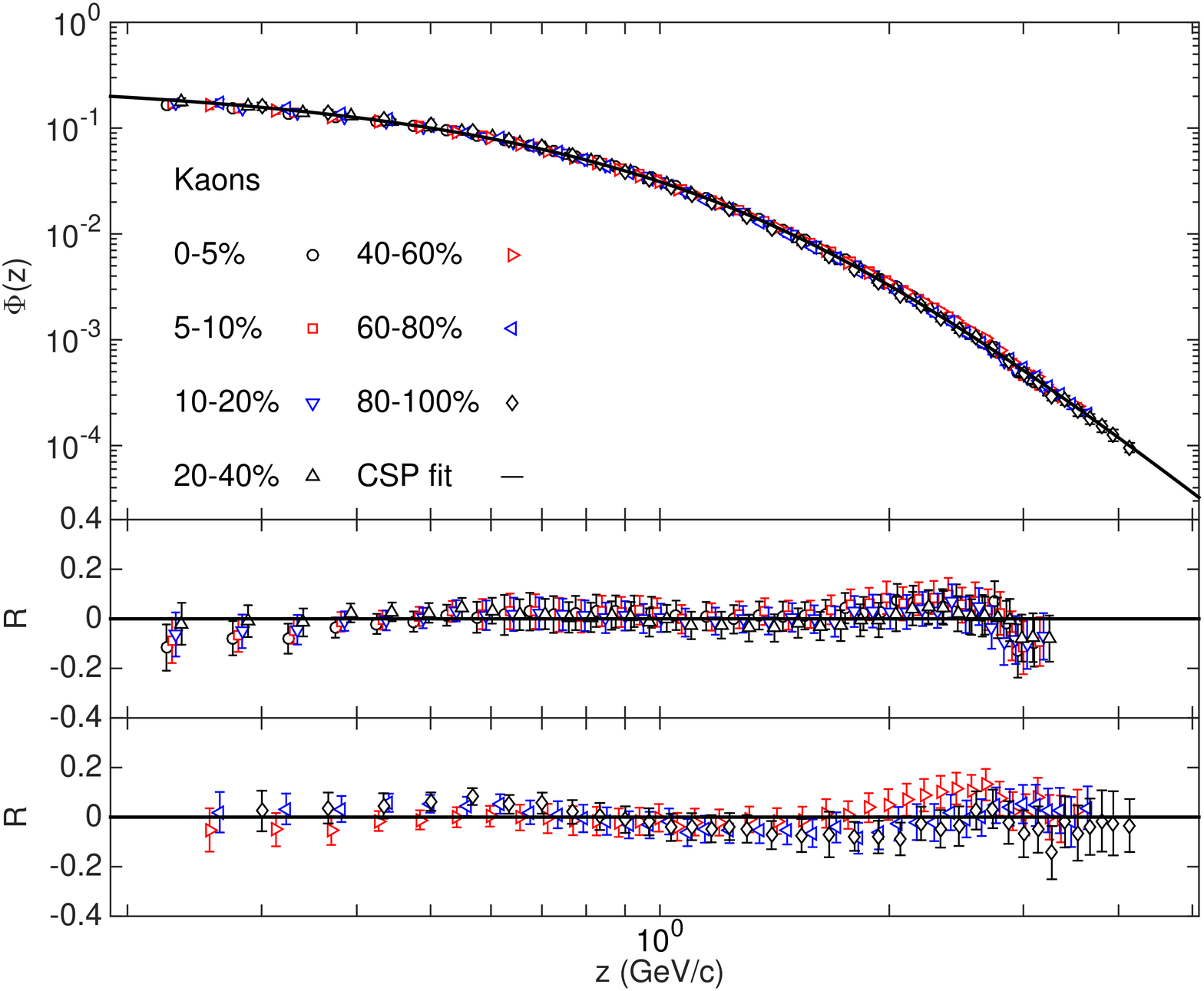}}\put(-36,102){(b)}
\caption{\label{fig:csp_pi_k} Upper panel in (a) ((b)): the scaling behaviour of the pion (kaon) $p_{\rm T}$ spectra presented in $z$. The black curve is CSP fit in eq. (\ref{eq:CPS_formula}) with parameters in second (third) column of Table \ref{tab:CSP_fit_parameters}. The data points are taken from ref. \cite{data_PI_K_P}. The middle (lower) panels in (a) and (b): the $R$ distributions at the 0-5$\%$, 5-10$\%$, 10-20$\%$ and 20-40$\%$ (40-60$\%$, 60-80$\%$ and 80-100$\%$) centrality classes.}
\end{figure}

\begin{figure}[H]
    \centering
\resizebox{0.42\textwidth}{!}{\includegraphics{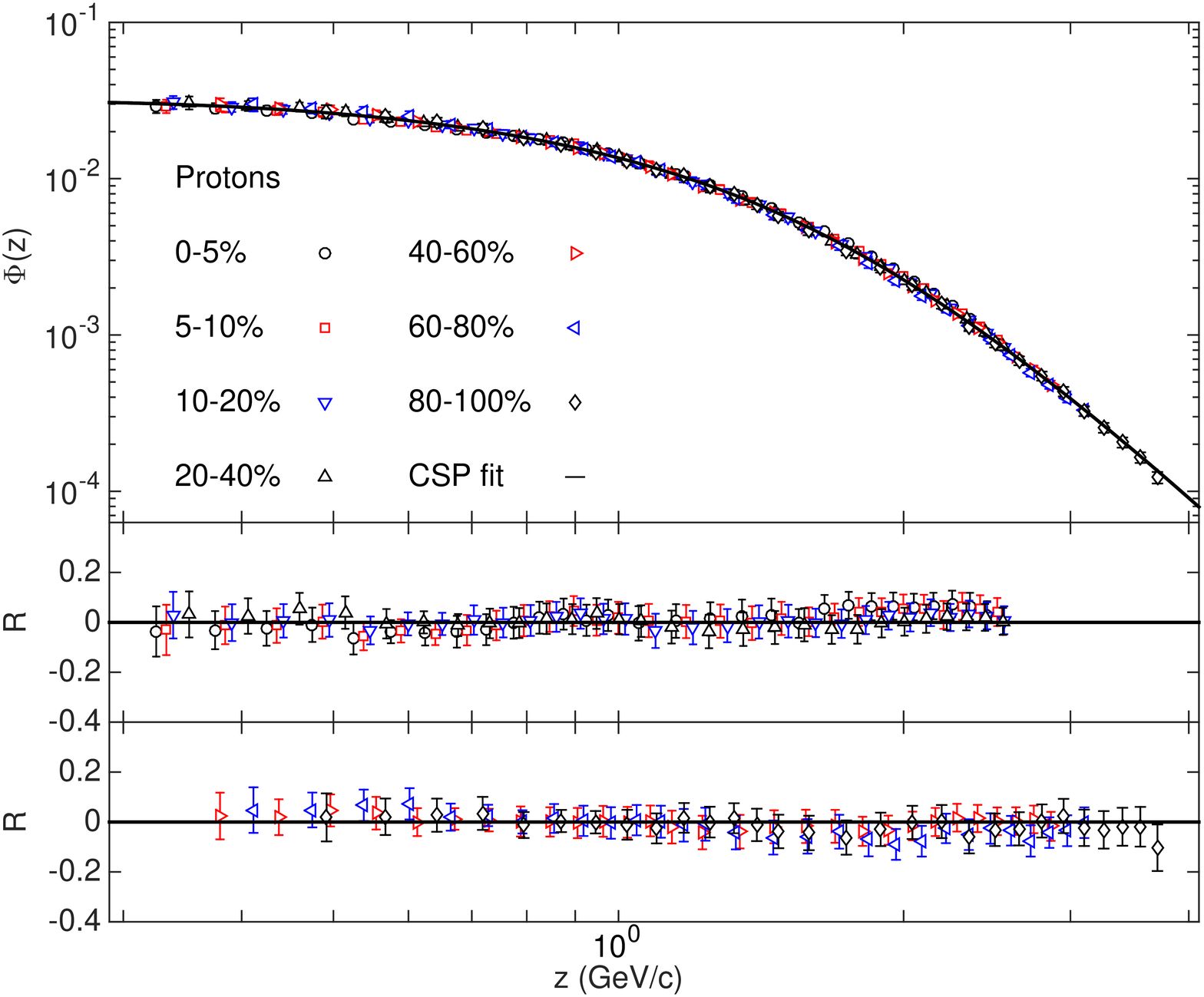}}\put(-36,102){(a)}
\resizebox{0.42\textwidth}{!}{\includegraphics{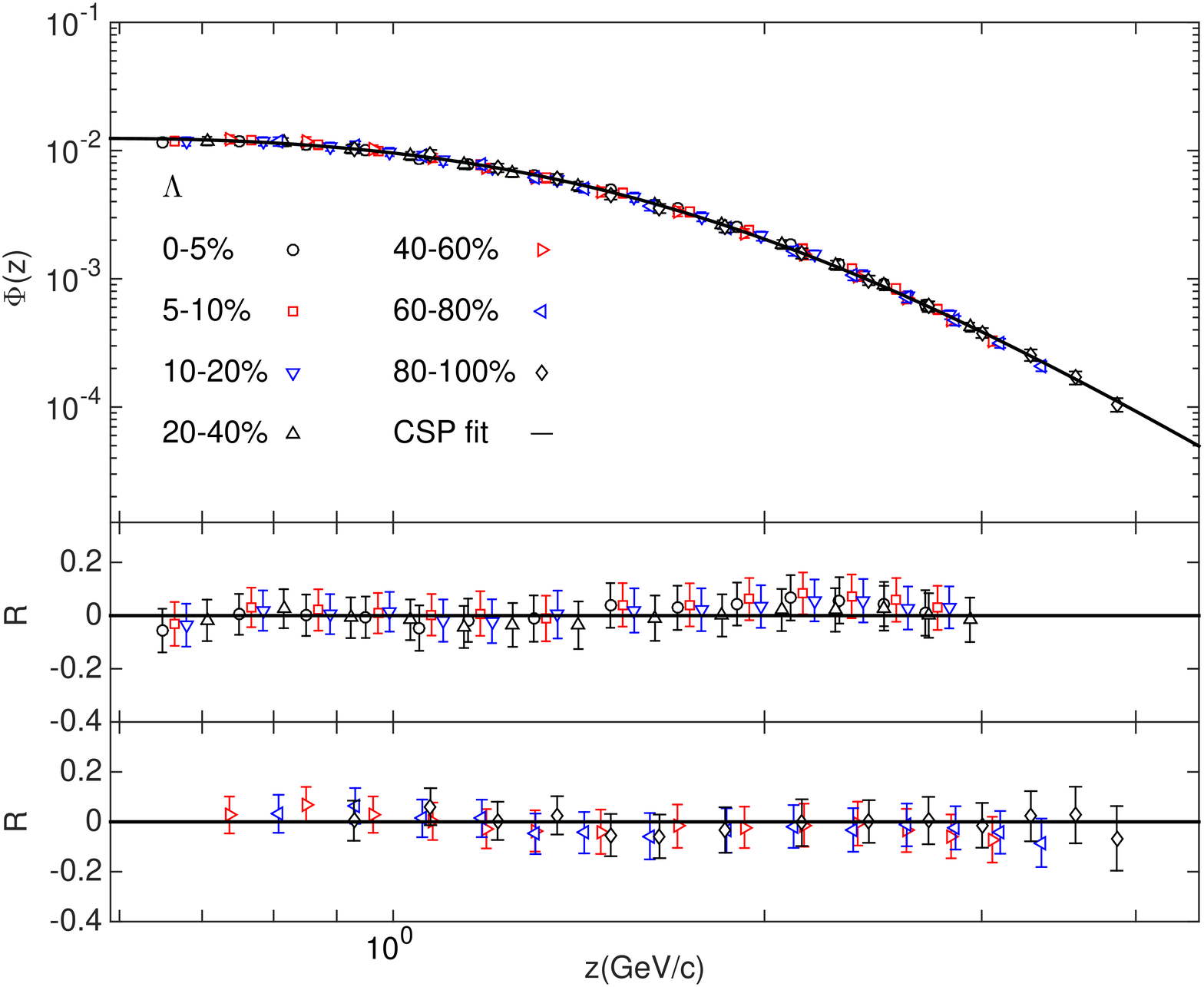}}\put(-36,102){(b)}
\caption{\label{fig:csp_p_lambda} Upper panel in (a) ((b)): the scaling behaviour of the proton ($\rm \Lambda$) spectra presented in $z$.  The black curve is CSP fit in eq. (\ref{eq:CPS_formula}) with parameters in the fourth (fifth) column of Table \ref{tab:CSP_fit_parameters}. The data points are taken from refs. \cite{data_PI_K_P, data_lambda_Ks0}. The middle (lower) panels in (a) and (b): the $R$ distributions at the 0-5$\%$, 5-10$\%$, 10-20$\%$ and 20-40$\%$ (40-60$\%$, 60-80$\%$ and 80-100$\%$) centrality classes.}
 \end{figure}

\begin{figure}[H]
    \centering
\resizebox{0.42\textwidth}{!}{\includegraphics{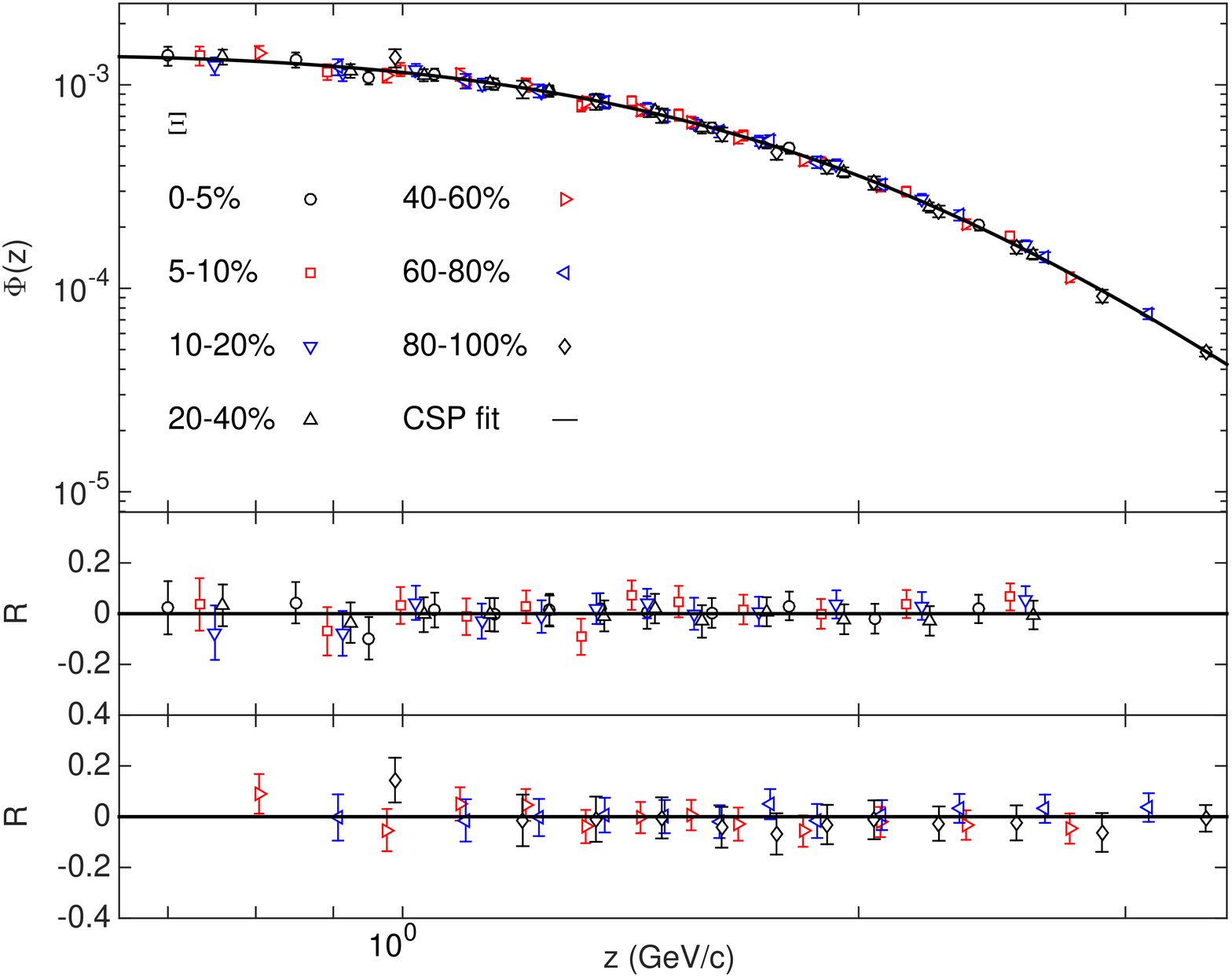}}\put(-36,102){(a)}
\resizebox{0.42\textwidth}{!}{\includegraphics{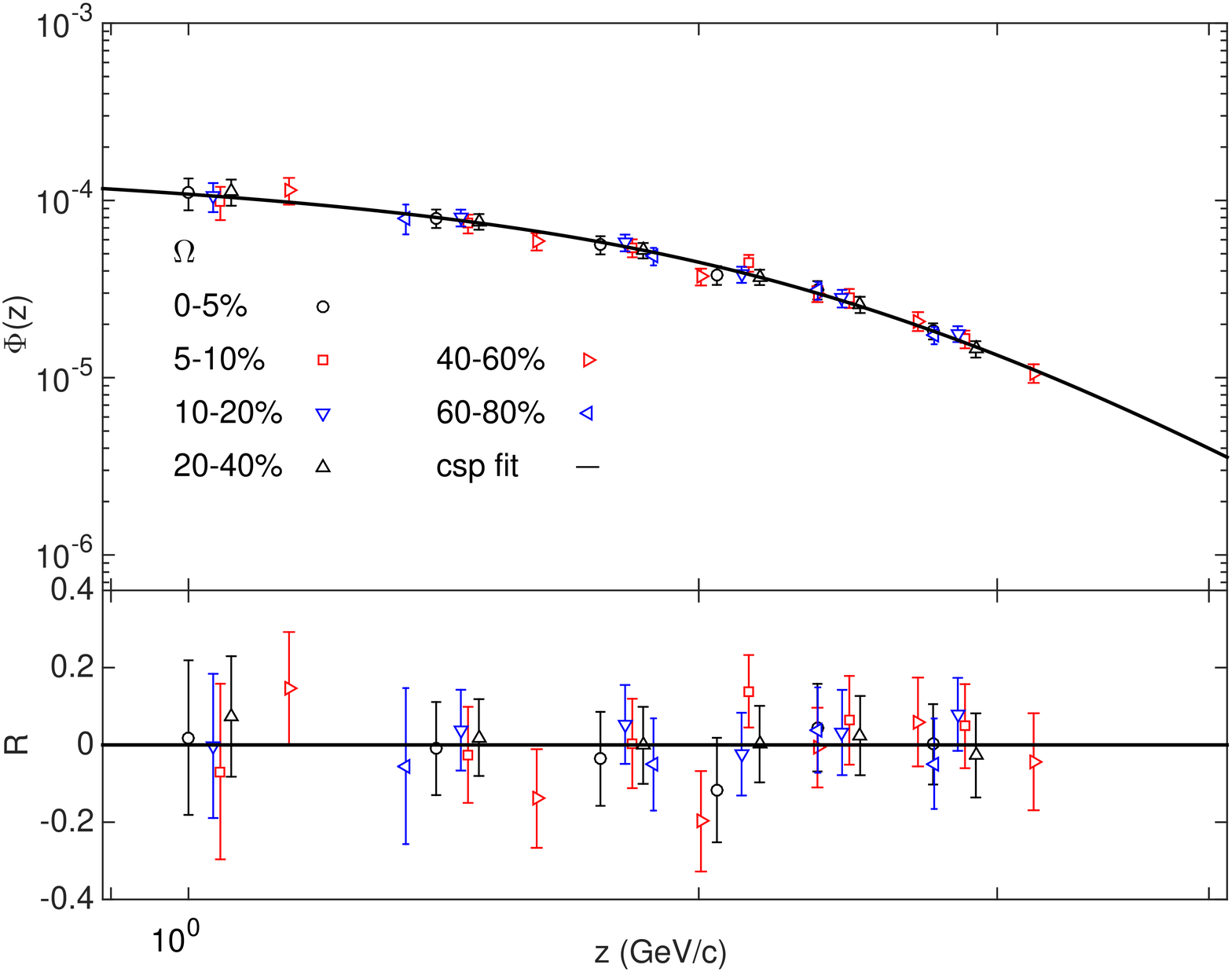}}\put(-36,102){(b)}
\caption{\label{fig:csp_XI_Omega} Upper panel in (a) ((b)): the scaling behaviour of the ${\rm \Xi}$ (${\rm \Omega}$) spectra presented in $z$. The black curve is CSP fit in eq. (\ref{eq:CPS_formula}) with parameters in the sixth (seventh) column of Table \ref{tab:CSP_fit_parameters}. The data points are taken from ref. \cite{data_XI_Omega}. The middle and lower panels in (a) ((b)): the $R$ distributions for the ${\rm \Xi}$ (${\rm \Omega}$)  spectra at the 0-5$\%$, 5-10$\%$, 10-20$\%$, 20-40$\%$, 40-60$\%$, 60-80$\%$ and 80-100$\%$ (0-5$\%$, 5-10$\%$, 10-20$\%$, 20-40$\%$, 40-60$\%$ and 60-80$\%$) centrality classes.}
\end{figure}

In order to check whether the CSP model can describe the scaling behaviour of the pion (kaon, proton, ${\rm \Lambda}$, ${\rm \Xi}$ and ${\rm \Omega}$) spectra in the low $p_{\rm T}$ region, eq. (\ref{eq:CPS_formula}) is fitted to the combination of the scaled data points with $p_{\rm{T}}\leq $ 3.9 (3.1, 2.5, 2.7, 2.4 and 2.8) GeV/c at different centrality classes using the least squares method. For the fits on the pion and kaon (proton, ${\rm \Lambda}$ and ${\rm \Xi}$) spectra, we found that the choice $\eta=0$ ($\theta=0$) is favored by the data. For the fit on the ${\rm \Omega}$ spectra, the choices of  $\alpha$ and $\theta$ both to be 0 are preferred by the data. In addition, as done in ref. \cite{frag_function_variable}, we have assumed that the mesons and baryons at different centralities are produced from clusters with the same size dispersion. With this assumption, we fixed the $\kappa$ values of kaons, protons, ${\rm \Lambda}$, ${\rm \Xi}$ and ${\rm \Omega}$ to the value of pions. The free parameters $C$, $\gamma$, $\kappa$, $\alpha$, $\eta$ and $\theta$ returned by the fits are tabulated in Table \ref{tab:CSP_fit_parameters}. From the table, we observe that the values of $\gamma$ for mesons are smaller than those for baryons, which means that the mean size of the cluster $\langle x\rangle$ for mesons is larger than that for baryons. This could be understood as follows. As shown in ref. \cite{data_lambda_Ks0}, $\langle p_{\rm T}\rangle$ is smaller for mesons than for baryons. As $x$ is proportional to $1/\langle p_{\rm T}^{2}\rangle$, $\langle x\rangle$  is greater for mesons than for baryons. The fit results for the pion, kaon, proton, ${\rm \Lambda}$, ${\rm \Xi}$ and ${\rm \Omega}$ $p_{\rm T}$ spectra are presented in the upper panels of Figs. \ref{fig:csp_pi_k} (a),  \ref{fig:csp_pi_k} (b),  \ref{fig:csp_p_lambda} (a), \ref{fig:csp_p_lambda} (b), \ref{fig:csp_XI_Omega} (a) and \ref{fig:csp_XI_Omega} (b)  respectively. The $R$ distributions are shown in the middle and lower panels of these figures. For the pion spectra, except for the first data point at the 80-100$\%$ centrality class, all the other data points at different centrality classes agree with the CSP fit within $20\%$. For the kaon, proton, ${\rm \Lambda}$, ${\rm \Xi}$ and ${\rm \Omega}$ spectra, all the data points at different centrality classes are consistent with the CSP fit within $20\%$.

\begin{table}[H]
  \caption{\label{tab:CSP_fit_parameters} $C$, $\gamma$, $\kappa$, $\alpha$, $\eta$ and $\theta$ returned by the CSP fits on the combination of the scaled pion, kaon, proton, ${\rm \Lambda}$, ${\rm \Xi}$ and ${\rm \Omega}$ $p_{\rm T}$ spectra at different centrality classes. The uncertainties quoted for pions are the errors returned from the fit. For kaons, protons, ${\rm \Lambda}$, ${\rm \Xi}$ and ${\rm \Omega}$, the uncertainties are determined by adding the errors returned from the fits and from the variation of $\kappa$ by $\pm 1\sigma$ in the fits in quadrature. The last row shows the reduced $\chi^{2}s$ of the fits.}
\begin{center}
\begin{tabular}{@{}ccccccc}
\toprule \textrm{\ } & \textrm{Pions}  & \textrm{Kaons} & \textrm{Protons} &\textrm{$\rm \Lambda$} &  \textrm{$\rm \Xi$}   &\textrm{$\rm \Omega$}\\
\hline
$C$                    & 6.01$\pm$0.41     & 16.51$\pm$2.24     & 0.09$\pm$0.02     & 0.28$\pm$0.08        & (94$\pm$17)$\times 10^{-4}$   & (190$\pm$25)$\times 10^{-6}$\\
$\gamma$               &7.51$\pm$0.73       & 9.49$\pm$0.68   & 16.47$\pm$1.89      &21.74$\pm$6.18        & 24.00$\pm$6.25               & 23.93$\pm$3.73 \\
$\kappa$               & 3.03$\pm$0.04      & 3.03(fixed)     &  3.03(fixed)        & 3.03(fixed)           & 3.03(fixed)                 & 3.03(fixed) \\
$\alpha$               & -0.24$\pm$0.02     & 1.45$\pm$0.05   & 0.29$\pm$0.07       & 1.26$\pm$0.39         &0.74$\pm$0.37                & 0(fixed)\\
$\eta$                 & 0 (fixed)          & 0 (fixed)          &3.12$\pm$0.40      & 4.61$\pm$1.77         &3.01$\pm$1.28               & 1.35$\pm$0.35\\
$\theta$               & -10.35$\pm$0.34    & -13.37$\pm$0.39    &0 (fixed)          & 0 (fixed)            & 0(fixed)                   & 0(fixed)\\
$\chi^{2}/ndf$         & 318.43/282         & 103.84/255        & 56.58/199          & 18.42/94             & 26.30/80                   & 12.48/33\\
\toprule
\end{tabular}
\end{center}
\end{table}

From the above statement, we observe that the CSP model can successfully describe the universal scaling of the pion, kaon, proton, ${\rm \Lambda}$, ${\rm \Xi}$ and ${\rm \Omega}$ spectra in the low $p_{\rm T}$ region. The universal scaling in the spectra of hadrons produced by clusters in the CSP model is guaranteed by the invariance of the cluster's size distribution $W(x)$ in eq. (\ref {eq:gamma_function}) and the fragmentation function $g(x, p_{\rm T})$ in eq. (\ref{eq:frag_function_form_1}) under the transformation $x \rightarrow x^{\prime} = \lambda x$, $\gamma \rightarrow \gamma^{\prime} = \gamma/\lambda$ and $p_{\rm T} \rightarrow p_{\rm T}^{\prime} = p_{\rm T}/\sqrt{\lambda}$, where $\lambda=\langle S_{n}/nS_{1} \rangle^{1/2}$, with the average taken over all the clusters decaying into hadrons \cite{string_perco_model_2}. Comparing the transformation $p_{\rm T}^{\prime} \rightarrow p_{\rm T}^{\prime}\sqrt{\lambda}$ in the CSP model with the linear transformation $p_{\rm T}\rightarrow p_{\rm T}/K$, we find that the scaling parameter $K$ is proportional to $\langle nS_{1}/S_{n} \rangle^{1/4}$. As described in ref. \cite{string_perco_model_1},  when going from peripheral to central collisions, the degree of string overlap $nS_{1}/S_{n}$ increases nonlinearly. Thus the scaling parameter $K$ should also increases with the centrality in a nonlinear trend. This is indeed what we observed  in Table \ref{tab:a_k_parameters}.

In order to determine the nonlinear trend, the $K$ values at different centrality classes is fitted with a function $K = a\langle N_{\rm{part}}\rangle^{b}$, where $a$ and $b$ are free parameters, $b$ represents the rate at which $\textrm {ln}{K}$ changes with $\textrm{ln}\langle N_{\rm{part}}\rangle$. In sect. \ref{sec:method}, the scaling parameter $K$ at the 0-5$\%$ centrality class is set to be 1.  In order to do the fit, we assign the relative error of $\langle p_{\rm T}\rangle$ as the uncertainty of $K$ at this centrality class. The $b$ values returned by the fits for pions, kaons, protons, ${\rm \Lambda}$, ${\rm \Xi}$ and ${\rm \Omega}$ are $0.150\pm0.017$, $0.166\pm0.013$, $0.202\pm0.013$, $0.197\pm0.014$, $0.199\pm0.012$ and $0.251\pm0.029$ respectively. The rate at which $\textrm {ln}{K}$ increases with $\textrm{ln}\langle N_{\rm{part}}\rangle$  is stronger for baryons than that for mesons. Moreover, for pions (kaons and protons), the rate is larger in p-Pb collisions than that in Pb-Pb collisions at 2.76 TeV in ref. \cite{Pb_Pb_pi_k_p}.

The different nonlinear trends for mesons and baryons in p-Pb collisions can also be explained by the CSP model as follows. Since $K$ is proportional to $\langle p_{\rm T}\rangle$, the ratio of the scaling parameter $K$ at non-central (5-10$\%$, 10-20$\%$, 20-40$\%$, 40-60$\%$, 60-80$\%$ and 80-100$\%$) collisions to that at central (0-5$\%$) collisions should be identical to the ratio between the values of  $\langle p_{\rm T}\rangle$ at non-central and central collisions.  In the CSP model, the $\langle p_{\rm T}\rangle$ value is evaluated as
\begin{eqnarray}
\langle p_{\rm T}\rangle=\frac{\int_{0}^{\infty}\int_{0}^{1/p^{2}_{\rm{T}}}W(x)g(x, p_{\rm T})p_{\rm T}^{2}dxdp_{\rm T}}{\int_{0}^{\infty}\int_{0}^{1/p^{2}_{\rm{T}}}W(x)g(x, p_{\rm T})p_{\rm T}dxdp_{\rm T}}.
\label{eq:p_T_CSP}
\end{eqnarray}
By substituting $W(x)$ in eq. (\ref{eq:gamma_function}) and $g(x, p_{\rm T})$ in eq. (\ref{eq:frag_function_form_1}) into the above equation, we obtain
\begin{eqnarray*}
\langle p_{\rm{T}}\rangle=\frac{\sqrt{\gamma}(\alpha+2)_{2}F_{1}(\alpha+3,-\theta,\alpha+\eta+4,-1)\Gamma(\kappa-3/2)} {(\alpha+\eta+3)_{2}F_{1}(\alpha+2,-\theta,\alpha+\eta+3,-1)\Gamma(\kappa-1)},
\label{eq:p_T_mean_r}
\end{eqnarray*}
where $_{2}F_{1}$ is the hypergeometric function. For the same hadron, as $g(x, p_{\rm T})$ are the same at different centrality classes, the parameters $\alpha$, $\eta$ and $\theta$ are identical at these centrality classes. In addition, as we have assumed that the dispersion of the cluster's size distributions for mesons and baryons at different centrality classes is identical, the values of $\kappa$ are the same at these centrality classes. Thus, the ratio between the values of $\langle p_{\rm T}\rangle$ at non-central and central collisions only relies on $\gamma$. The $\gamma$ values at different centrality classes are determined by fitting the spectra of the pion (kaon, proton, ${\rm \Lambda}$, ${\rm \Xi}$ and ${\rm \Omega}$ ) in the low $p_{\rm T}$ region with $p_{\rm{T}}\leq $ 3.9 (3.1, 2.5, 2.7, 2.4 and 2.8) GeV/c to eq. (\ref{eq:CPS_formula}) with $\alpha$, $\eta$, $\theta$ and $\kappa$ fixed to their center values in Table \ref{tab:CSP_fit_parameters}. They are listed in Table \ref{tab:r_k_parameters}. Also tabulated in the table are the reduced $\chi^{2}$s. For the fit on the pion spectrum at the 80-100$\%$ centrality class, the reduced $\chi^{2}$ is large, as there is an deviation between the fitted curve and the data points in the region with $p_{\rm T}\leq$ 0.2 GeV/c. The uncertainties of $\gamma$ are determined by adding the errors returned from the fits and from the variation of $\alpha$, $\eta$, $\theta$ and $\kappa$ in turn by $\pm1\sigma$ in the fits in quadrature. From the table, for a species of hadrons, within uncertainties we see that with the increase of centralities $\gamma$ increases, which implies that the mean size of the cluster $\langle x\rangle$ decreases. This is due to the reason that the mean $p_{\rm T}$ increases with centralities and $x$ is proportional to $1/\langle p_{\rm T}^{2}\rangle$ \cite{data_lambda_Ks0}. With these $\gamma$ values, we can calculate the ratios between the values of $\langle p_{\rm T}\rangle$ at non-central and central collisions. The results are listed in Table \ref{tab:PT_mean}. For different hadron species, when comparing these ratios with the scaling parameters $K$ at non-central collisions in Table \ref{tab:a_k_parameters}, we find that they are in agreement within the uncertainties. Therefore, the universal scaling  can be quantitatively understood by the CSP model at the same time.

\begin{table}[H]
  \caption{\label{tab:r_k_parameters} $\gamma$ of the CSP fits on the pion, kaon, proton, ${\rm \Lambda}$, ${\rm \Xi}$ and ${\rm \Omega}$ spectra at different centrality classes. The uncertainties quoted are determined by adding the errors returned from the fits and from the variation of $\alpha$, $\eta$, $\theta$ and $\kappa$ in turn by $\pm1\sigma$ in the fits in quadrature. The last column shows the reduced $\chi^{2}$ for the fits.}
\begin{center}
\begin{tabular}{@{}cccc}
\toprule {\ } & {Centrality}  & \textrm{$\gamma$} &\textrm{$\chi^{2}$/dof}\\
\hline
\multirow {7}{*}{Pions}
&0-5$\%$&       7.68$\pm$0.62&     13.71/39\\
&5-10$\%$&      7.59$\pm$0.61&     7.22/39\\
&10-20$\%$&     7.38$\pm$0.60&     4.91/39\\
&20-40$\%$&     7.00$\pm$0.56&     4.91/39\\
&40-60$\%$&     6.40$\pm$0.51&     7.38/39\\
&60-80$\%$&     5.61$\pm$0.44&     18.75/39\\
&80-100$\%$&    4.18$\pm$0.36&     122.07/39\\
\hline
\multirow {7}{*}{ Kaons}
&0-5$\%$&         9.76$\pm$0.82&      12.02/35\\
&5-10$\%$&        9.52$\pm$0.78&      13.58/35\\
&10-20$\%$&       9.05$\pm$0.72&      9.39/35\\
&20-40$\%$&       8.52$\pm$0.72&      5.37/35\\
&40-60$\%$&       7.64$\pm$0.61&      6.97/35\\
&60-80$\%$&       6.71$\pm$0.53&      14.18/35\\
&80-100$\%$&      5.07$\pm$0.39&      11.21/35\\
\hline
\multirow {7}{*}{Protons}
&0-5$\%$&         17.64$\pm$2.57&     3.38/27\\
&5-10$\%$&        16.33$\pm$2.35&     2.97/27\\
&10-20$\%$&       15.34$\pm$2.17&     2.14/27\\
& 20-40$\%$&      13.99$\pm$1.92&     3.44/27\\
&40-60$\%$&       11.98$\pm$1.61&     2.89/27\\
&60-80$\%$&       9.85$\pm$1.28&      5.27/27\\
&80-100$\%$&      7.07$\pm$0.89&      3.79/27\\
\hline
\multirow {7}{*}{${\rm \Lambda}$}
&0-5$\%$&         22.6$\pm$10.9&      1.12/12\\
&5-10$\%$&        21.51$\pm$10.07&    0.96/12\\
&10-20$\%$&       20.44$\pm$9.24&     0.66/12\\
&20-40$\%$&       18.61$\pm$8.53&     0.97/12\\
&40-60$\%$&       16.33$\pm$7.09&     1.23/12\\
&60-80$\%$&       13.6$\pm$5.8&       1.58/12\\
&80-100$\%$&      10.52$\pm$4.40&     2.03/12\\
\hline
\multirow {7}{*}{${\rm \Xi}$}
&0-5$\%$&        23.49$\pm$11.89&      3.38/11\\
&5-10$\%$&       22.28$\pm$10.86&      5.52/11\\
&10-20$\%$&      21.73$\pm$9.67&       2.12/11\\
& 20-40$\%$&     19.96$\pm$9.13&       1.04/11\\
&40-60$\%$&      17.39$\pm$7.10&       3.05/11\\
&60-80$\%$&      14.81$\pm$6.05&       1.18/11\\
&80-100$\%$&     12.11$\pm$4.49&       3.67/11\\
\hline
\multirow {6}{*}{$\rm \Omega$}
&0-5$\%$&        22.55$\pm$3.76&      1.85/5\\
&5-10$\%$&       23.85$\pm$3.76&      1.27/5\\
&10-20$\%$&      21.8$\pm$3.6&       1.08/5\\
& 20-40$\%$&     19.99$\pm$3.93&      0.37/5\\
&40-60$\%$&      17.61$\pm$3.34&      5.69/5\\
&60-80$\%$&      13.43$\pm$2.59&      1.50/5\\
\toprule
\end{tabular}
\end{center}
\end{table}

\begin{table}[H]
  \caption{\label{tab:PT_mean} The ratio between the values of $\langle p_{\rm T}\rangle$ at non-central and central collisions for pions, kaons, protons, ${\rm \Lambda}$, ${\rm \Xi}$ and ${\rm \Omega}$. The uncertainties quoted are due to the errors of $\gamma$.}
\begin{center}
\begin{tabular}{@{}ccccccc}
\toprule \textrm{Centrality}&\textrm{Pions}&\textrm{Kaons}&\textrm{Protons} &\textrm{$\rm \Lambda$}&\textrm{$\rm \Xi$}&\textrm{$\rm \Omega$}\\
\hline
5-10$\%$&   0.99$\pm$0.06& 0.99$\pm$0.06&  0.97$\pm$0.10& 0.98$\pm$0.33&  0.97$\pm$0.34&  1.03$\pm$0.12\\
10-20$\%$&  0.98$\pm$0.06& 0.96$\pm$0.06&  0.94$\pm$0.10&  0.95$\pm$0.31&  0.96$\pm$0.32&  0.98$\pm$0.12\\
20-40$\%$&  0.95$\pm$0.05& 0.93$\pm$0.06&  0.90$\pm$0.09&  0.91$\pm$0.30&  0.92$\pm$0.31&  0.94$\pm$0.12\\
40-60$\%$&  0.91$\pm$0.05& 0.88$\pm$0.05&  0.83$\pm$0.08&   0.85$\pm$0.28&  0.86$\pm$0.28&  0.88$\pm$0.11\\
60-80$\%$&  0.85$\pm$0.05& 0.83$\pm$0.05&  0.75$\pm$0.07&   0.78$\pm$0.25&  0.79$\pm$0.26&  0.77$\pm$0.10\\
80-100$\%$& 0.74$\pm$0.04& 0.72$\pm$0.04&  0.64$\pm$0.06&   0.68$\pm$0.22&  0.72$\pm$0.23&       $-$\\
\toprule
\end{tabular}
\end{center}
\end{table}

\section{Conclusions}\label{sec:conclusion}
In this paper, we have a systematic study on the scaling property of the meson and baryon spectra at different centrality classes in p-Pb collisions at 5.02 TeV. When presented in terms of a suitable variable, $z=p_{\rm T}/K$, we found that the pion (kaon, proton, ${\rm \Lambda}$, ${\rm \Xi}$ and ${\rm \Omega}$) spectra exhibit a universal scaling independent of the centrality class in the low $p_{\rm T}$ region with $p_{\rm{T}}\leq$ 3.9 (3.1, 2.5, 2.7, 2.4 and 2.8)  GeV/c. The scaling parameter $K$ depends on the centrality class and the rates at which ln$K$ increases with $\textrm{ln}\langle N_{\rm{part}}\rangle$ for pions and kaons are smaller than those for protons, ${\rm \Lambda}$, ${\rm \Xi}$ and ${\rm \Omega}$. In the high $p_{\rm T}$ region, there is a deviation from the universal scaling going from central to peripheral collisions. The more peripheral the collisions are, the more obvious the violation of the scaling is. In the framework of the CSP model, we argue that the pion, kaon, proton, ${\rm \Lambda}$, ${\rm \Xi}$ and ${\rm \Omega}$ are produced by the fragmentation of clusters formed by strings overlapping in the transverse plane with the same size dispersion but with different mean size. The mean size of clusters for baryons is smaller than for mesons. For the same hadrons at different centrality classes, the mean size of clusters decreases with the increase of centrality. The cluster's fragmentation functions are different for different hadrons, while they are the same for a species of hadrons at different centrality classes.  The universal scaling of the meson and baryon spectra in the low $p_{\rm T}$ region can be quantitatively understood with the CSP model at the same time.

\section*{Acknowledgements}
 This work was supported by the Fundamental Research Funds for the Central Universities of China under Grant No. GK201803013, by the Scientific Research Foundation for the Returned Overseas Chinese Scholars, State Education Ministry, by Natural Science Basic Research Plan in Shaanxi Province of China (program No. 2017JM1040) and by the National Natural Science Foundation of China under Grant Nos. 11447024 and 11505108.


\end{document}